%% file: main.tex
\documentclass[11pt]{article}

\newcommand{\blind}{1}

\input{"./preamble/packages"}
\input{"./preamble/definition"}
\input{"./preamble/acronym"}

\newcommand{\mak}[1]{{\color{blue} \textbf{mak:} #1}}
\newcommand{\bac}[1]{{\color{red} \textbf{bac:} #1}}
\newcommand{\jzl}[1]{{\color{brown} (\textbf{jzl:} #1)}}

\newcommand{\PM}{PM$_{2.5} $}

\begin{document}

\def\spacingset#1{\renewcommand{\baselinestretch}%
{#1}\small\normalsize} \spacingset{1}

\if1\blind
{
  \title{\bf Spatially Adaptive Ensemble Learning with Calibrated Predictive Uncertainty}
    \renewcommand{\Authfont}{\normalfont\large}
    \renewcommand{\Affilfont}{\normalfont\small}
    \renewcommand{\Authsep}{\quad}
    \renewcommand{\Authands}{\quad}
    \setlength{\affilsep}{0.6em}
    \author[1,2]{Yanran Li}
    \author[1,3]{Jeremiah Zhe Liu}
    
    \author[2]{John Paisley}
    \author[2]{Marianthi-Anna Kioumourtzoglou}
    \author[1]{Brent A. Coull}
    \makeatletter
    \@namedef{@sep4}{\protect\\[3pt]}
    \renewcommand{\AB@affilsepx}{\quad\protect\Affilfont}
    \makeatother
    \affil[1]{Harvard University}
    \affil[2]{Columbia University}
    \affil[3]{Meta}
    \affil[4]{Brown University}
    \date{}                     
    \maketitle
} \fi

\if0\blind
{
  \bigskip
  \bigskip
  \bigskip
  \begin{center}
    {\LARGE\bf Spatially Adaptive Ensemble Learning with Calibrated Predictive Uncertainty: Application to Ambient Air Pollution Prediction Modeling}
\end{center}
  \medskip
} \fi

\bigskip
\input{"./sections/abstract_v1"}

\noindent%
\vfill

\input{"./sections/intro_v5_nofig"}
\input{"./sections/model"}
\input{"./sections/inference_v1_nofig"}

\input{"./sections/simu_nofig"}
\input{"./sections/application_nofig"}
\input{"./sections/conclusion"}

\section*{Acknowledgments}
This publication was made possible by USEPA grant RD-83587201. Its contents are solely the responsibility of the grantee and do not necessarily represent the official views of the USEPA. Further, USEPA does not endorse the purchase of any commercial products or services mentioned in the publication. Funding was also provided by NIEHS grants R01 ES030616, P30 ES000002, and P30 ES009089.

\clearpage
\bibliographystyle{abbrv}
\bibliography{report,additional}

\clearpage
\section*{Tables and Figures}

\input{"./sections/table"}

\input{"./sections/figure"}

\clearpage
\section*{Supplementary Material}
\appendix
\beginsupplement
\input{"./sections/appendix_v1"}
\clearpage
\section{Tables and Figures}
\input{"./sections/table_app"}
\clearpage
\input{"./sections/figure_app"}


\end{document}

%% file: preamble/packages.tex
\usepackage[a4paper]{geometry}

\usepackage{graphicx}
\usepackage{enumerate}
\usepackage{natbib}
\usepackage{url} 

\usepackage{graphicx,psfrag,epsf}
\usepackage{enumerate}
\usepackage{natbib}
\usepackage{authblk}

\usepackage[T1]{fontenc}    
\usepackage{amsfonts}       

\usepackage{adjustbox}
\usepackage{multirow}
\usepackage{etoolbox}
\usepackage[caption=false]{subfig}
\usepackage{amsmath}
\usepackage{amssymb}
\usepackage{mathtools}
\usepackage{amsfonts}
\usepackage{mathptmx}
\usepackage{textgreek}
\usepackage{amsthm}

\usepackage{makecell}

\usepackage
[acronym,nowarn,section,nonumberlist]{glossaries}
\glsdisablehyper{}

\usepackage{color} 
\usepackage{xcolor}
\usepackage{"macro/GrandMacros"}
\usepackage{"macro/Macro_BIO235"}

\usepackage{comment}

%% file: preamble/definition.tex
\newtheoremstyle{break}
  {\topsep}{\topsep}%
  {\itshape}{}%
  {\bfseries}{}%
  {\newline}{}%
\theoremstyle{break}

\newtheorem{theorem}{Theorem}

\makeatletter
\patchcmd{\@makecaption}
  {\parbox}
  {\advance\@tempdima-\fontdimen2} 
  {}{}
\makeatother  

\newcommand{\beginsupplement}{%
        \setcounter{table}{0}
        \renewcommand{\thetable}{S\arabic{table}}%
        \setcounter{figure}{0}
        \renewcommand{\thefigure}{S\arabic{figure}}%
     }

%% file: preamble/acronym.tex
\newacronym{UQ}{UQ}{uncertainty quantification}

\newacronym{MAP}{MAP}{maximum a posterior}

\newacronym{CDF}{CDF}{cumulative distribution function}
\newacronym{PDF}{PDF}{probability density function}
\newacronym{MGF}{MGF}{moment generating function}

\newacronym{GLM}{GLM}{generalized linear model}
\newacronym{GAM}{GAM}{generalized additive model}
\newacronym{MLE}{MLE}{maximum likelihood estimation}

\newacronym{HMC}{HMC}{Hamiltonian Monte Carlo}
\newacronym{VI}{VI}{Variational Inference}
\newacronym{NUTS}{NUTS}{No-U-Turn Sampler}
\newacronym{OOD}{OOD}{out-of-distribution}
\newacronym{SGLD}{SGLD}{Stochastic-gradient Laungevin Dynamics}
\newacronym{SGHMC}{SGHMC}{Stochastic-gradient Hamiltonian Monte Carlo}

\newacronym{CRPS}{CRPS}{\textit{continous ranked probability score}}
\newacronym{RMSE}{RMSE}{root mean squared error}
\newacronym{MSE}{MSE}{mean squared error}
\newacronym{ECE}{ECE}{expected coverage error}
\newacronym{NLL}{NLL}{negative log likelihood}

\newacronym{KL}{KL}{Kullback-Leibler}
\newacronym{CvM}{CvM}{Cram\'{e}r von Mises}
\newacronym{ELBO}{ELBO}{\emph{evidence lower bound}}

\newacronym{BMA}{BMA}{Bayesian model averaging}
\newacronym{BNE}{BNE}{\textit{Adaptive Bayesian Nonparametric Ensemble}}
\newacronym{CVI}{Calibrated VI}{\textbf{Calibrated Variational Inference}}
\newacronym{EnKF}{EnKF}{Ensemble Kalman Filters}

\newacronym{RKHS}{RKHS}{reproducing kernel Hilbert space}
\newacronym{RBF}{RBF}{radial basis function}
\newacronym{CI}{CI}{credible intervals}

\newacronym{GP}{GP}{Gaussian process}
\newacronym{CGP}{CGP}{\textit{constrained Gaussian process}}
\newacronym{HEX-GP}{\textbf{HEX-GP}}{\textit{\underline{H}eteroskedastic \underline{Ex}ponentially-modified \underline{G}aussian \underline{p}rocess}}
\newacronym{EMG}{EMG}{\textit{Exponentially-modified Gaussian}}

%% file: sections/abstract_v1.tex
\begin{abstract}

Air pollution epidemiology continues to be a central area of investigation in the environmental health sciences. Current standard practice in the field is to employ
sophisticated spatio-temporal prediction models to estimate exposure to air pollution both at the individual and community levels.  A key methodological issue that arises in this context is uncertainty quantification of the resulting exposure estimates. An emerging technique in air pollution exposure assessment is to ensemble several competing exposure prediction models. Ensemble learning is a mainstay in modern data science. Conventional practice assigns to base models a set of deterministic, constant model weights that (1) do not fully account for the individual models' varying accuracy across sub-regions in the feature space and (2) do not provide uncertainty estimates for the ensemble prediction. The first of these shortcomings can yield suboptimal predictions in certain sub-regions of the study, while the second precludes the assessment of prediction reliability and propagation of uncertainty in downstream health effects analyses.  
We introduce a new Bayesian nonparametric ensemble framework that uses a dependent random measure to adaptively combine models based on their predictive accuracy in the feature space and also nonparametrically models the ensemble's predictive cumulative density function (CDF) so that the model's quantification of the predictive uncertainty is consistent with observed data. We show that the proposed method is asymptotically consistent for uncertainty quantification (i.e. CDF estimation) and can improve upon the predictive performance of an  ensemble model with na\"ive distribution assumptions when the data distribution is complex.
We apply the method to data simulated from one- and two-dimensional complex nonlinear regression models, and generate a spatial prediction model and estimate the associated prediction uncertainties for fine particle levels in eastern New England, USA.
\end{abstract}

%% file: sections/intro_v5_nofig.tex
\section{Introduction}

Ambient air pollution is a well-recognized global environmental threat to human health. In 2015, exposure to \PM\ (particles with aerodynamic diameter $\le$ 2.5 \textmu m) was the fifth-ranking mortality risk factor, causing 4.2 million deaths globally \citep{cohen_estimates_2017}, and the sixth-ranking factor for disability-adjusted life-years (DALYs), contributing to 103.1 million DALYs \citep{forouzanfar_global_2016}. These findings have been confirmed by numerous studies, including three US-wide studies of more than 60 million Medicare beneficiaries that reported significant \PM\ impacts on mortality, even at levels below the current national standards \citep{di_association_2017, di_air_2017, wu2020evaluating}. 

Despite the long-standing scientific consensus on the harmful effects of \PM\ exposures on the public's health, new questions have been raised about the methodology used to generate this evidence. A {\em Nature} editorial recently stated this ``scientific consensus is now under attack'', both in the United Stated and other countries \citep{tollefson_air_2019}. One major criticism levied against air pollution epidemiological studies is that health effect analyses treat estimated ambient air pollution levels as known \citep{u.s._environmental_protection_agency_casac_2019}.  Because it is not feasible---neither cost-wise nor logistically---to ask millions of study participants to constantly wear personal air pollution monitors for long periods of time, studies must rely on estimates of ambient air pollution concentration levels across an area of study.    

The current standard practice in the field is to employ sophisticated spatio-temporal prediction models to predict air pollution exposures at residential addresses of study participants, even at locations where the monitoring data are sparse or non-existent \citep{di_assessing_2016, di_hybrid_2016,  eeftens_development_2012, kim_historical_2017, kloog_new_2014, kloog_assessing_2011,  kloog_incorporating_2012,van_donkelaar_high-resolution_2015,wang_evaluation_2013,yanosky_spatio-temporal_2014}. However, due to the difference in data sources and in modeling assumptions, the accuracy of the predictions generated by different air pollution exposure models can differ in space and especially at locations that are far away from the air pollution monitors.  
As a result, health effect estimates obtained using exposure estimates from different exposure models can vary, making it important to consider model uncertainty associated with exposure assignment. For example, a recent study of the association between long-term \PM\ and coronary artery disease compared different exposure models and found significant effect modification by urban vs. rural residence only for some of the exposure models used, but not all \citep{mcguinn_fine_2017}. For the purposes of developing regulatory policy, such model uncertainty is often 
assessed informally but not quantified rigorously. 




In addition to sensitivity of conclusions to choice of exposure model, existing work on methods that properly account for the prediction error when the estimates are used as a covariate in subsequent health effects analyses has shown that failure to adjust for such uncertainty can adversely affect inference on these effects \citep{gryparis_measurement_2008, szpiro_does_2011, alexeeff_spatial_2016}.
All of this existing work aims to propagate statistical uncertainty associated with predictions of a single exposure model, which underestimates uncertainty associated with the selection of the exposure model.  Further, Alexeeff et al. (2017) showed that use of exposure estimates without accounting for spatially varying prediction error can yield biased health effect estimates.   Thus,  we seek to quantify both intra- and inter-model uncertainty and how it varies in either space and/or time, for use in future health effects analyses. 

Consequently, to establish rigorous scientific evidence on the health effects of air pollution, 
it is desirable to produce an ensemble framework for exposure estimation that accomplishes two primary objectives. The 
first is that it flexibly integrates information from available base exposure models in order to yield the most accurate estimates possible.   This objective requires that the ensemble model should allow \textbf{\textit{spatial adaptivity}} in its ensemble weights, which is necessary because some exposure models outperform competitors in some areas but not others \citep{kelly2020examining, jin2019comparison}. For instance, one model might out-perform another in urban but not rural settings, or near roadways or other microenvironments, or in high but not low concentration settings.  

The resulting ensemble must also enable valid quantification of prediction uncertainty.  This goal requires that an ensemble model achieves \textbf{\textit{calibrated estimation}} of its predictive uncertainty. 
In intuitive terms, this means a model's uncertainty estimate reliably reflects the discrepancy between the model prediction and the true observation, such that, for instance,  the model's $95\%$ predictive intervals contain new observations $95\%$ of the time \citep{gneiting_probabilistic_2007}.
Importantly, to properly quantify predictive uncertainty, the model should recognize different sources of uncertainty that arise in the modeling process.
In natural science applications, two distinct types of uncertainties exist: \textit{aleatoric uncertainty} and \textit{epistemic uncertainty} \citep{kiureghian_aleatory_2009, sullivan_introduction_2015, hullermeier2021aleatoric, matthies2007quantifying}. Aleatoric uncertainty arises due to stochastic variability inherent in the probability distribution of the data, while the epistemic uncertainty arises due to our lack knowledge of the data-generating mechanism (e.g., predict $y$ at a location $\bx$ not well covered by training data). A model's epistemic uncertainty can be reduced by collecting more data, whereas aleatoric uncertainty is irreducible since it is inherent to the data generating mechanism. Consequently, in regions that contain training data, a model's aleatoric uncertainty should accurately estimate the data-generating distribution by flexibly capturing the stochastic pattern in the data, while in regions in which training data is unavailable, the model's epistemic uncertainty should increase in order to capture the model's lack of confidence in the resulting predictions. 
The quality of a Bayesian model's uncertainty estimation can be measured rigorously using \textit{proper scoring rules} (e.g., the posterior predictive negative log likelihood  
under a Gaussian model)\cite{dawid_well-calibrated_1982, murphy_verification_1967, gneiting_strictly_2007, seidenfeld_calibration_1985}. To this end, the model's ability to provide a valid quantification of the aleatoric and epistemic uncertainty is essential for ensuring good performance, especially in \gls{OOD} regions not well-covered by the training data.

\subsection{Contribution to the Model-based Ensemble Literature}
In recent years, several ensemble methods have been developed in the environmental literature for air pollution prediction and climate forecasts. Examples include bootstrap aggregation of regression trees or neural networks \citep{li_ensemble_2017, hu_estimating_2017, wang_deep_2018}, hierarchical models with sophisticated covariance structure \citep{shaddick_data_2016, li_constrained_2017}, and additive combinations of multiple machine learning predictors \citep{di_ensemble-based_2019, xiao_ensemble_2018}. 
Bayesian variants of the ensemble models also exist, with examples including those of 
Smith et al. and Tebaldi et al. (\cite{tebaldi_quantifying_2005, smith_bayesian_2009}) that place a structured prior on the weights of an additive ensemble, and Gneiting and Weather  (\citep{gneiting_weather_2005}), Raftery (\citep{raftery_using_2005}), and Murray et al (\citep{murray2019bayesian}) that perform \gls{BMA} on the base models' predictive distributions. 
However,  these existing approaches either do not allow spatial adaptivity in the ensemble weights or assume that the true data generating model lies strictly within the convex combination of the base model predictions (e.g., the $\Msc$-closed assumption made by \gls{BMA})\citep{clydec_bayesian_2013}. Moreoever, 
some make parametric distributional assumptions for the residual errors in the ensemble \citep{murray2019bayesian}. 

In this work, we present a principled Bayesian approach to ensemble learning that achieves both a \textbf{\textit{spatially adaptive ensemble}} and \textbf{\textit{calibrated predictive uncertainty}}. That is, the approach adaptively combines predictions from multiple spatio-temporal processes, addresses model biases in the predictive distribution assumptions using Bayesian nonparametric machinery, and achieves a complete quantification of both the aleatoric and epistemic uncertainty. Specifically, we model the ensemble weights as a spatially dependent random measure, and then couple the ensemble model's prediction function with a residual process that mitigates the systematic bias shared by the base models. 
We also model the ensemble's distribution function semi-parametrically as a random function, carefully specifying its priors so that the posterior predictive distribution flexibly captures the data's empirical distribution while 
maintaining statistical efficiency via shrinkage mechanisms.
We term our method \gls{BNE}. From the perspective of statistical modeling, \gls{BNE} can be understood as a $\Msc$-open generalization of \gls{BMA}, where instead of restricting the model's distribution function to be a convex combination of the base model predictions, \gls{BNE} uses Bayesian nonparametric machinery to generate a large model space 
for the distribution function that is \textit{a priori} centered at the predictions from  the original ensemble. As a result, even when the base models are misspecified, \gls{BNE} is able to achieve 
improved predictive accuracy and calibrated predictive intervals.


\subsection{The Data: PM$_{2.5}$ in Eastern New England, USA}
\label{sec:data}

Eastern New England, USA is an area in which many different air pollution cohort studies are on-going, including cohorts of children (Project Viva), middle aged populations (Framingham Heart Study), and the elderly
(the Normative Aging Study), among others \citep{jhun_synthesis_2019, fleisch_prenatal_2017, dorans_residential_2016}.  For this case study, for our ensemble model we choose to model   annual PM$_{2.5}$ levels  in eastern New England during 2011.  For our ensembling framework we use data from three published PM$_{2.5}$ prediction models \citep{kloog_new_2014, van_donkelaar_high-resolution_2015, di_hybrid_2016}.  The main criterion for choice of the base models is the fact that the resulting predictions are freely available, ensuring the reproducibility of our work. 
Although all three models  reported roughly similar predictive accuracy in the broader study areas for which they were developed, they yielded different predictions in eastern New England, the focus of our case study.  This highlights 
the importance of adaptive weights in any ensemble model based on these predictions.  We provide some detail on the data and algorithms that yielded each set of predictions here.

The top row of Figure 1 shows estimated 2011 annual PM$_{2.5}$ concentration levels generated by the three models used in this work.  We use naming conventions for the models based on initials of the first author and year of the publication describing the each model.  The first model (MH2021; Figure \ref{fig:pm25_base}, Top Left) used geographically weighted regression (GWR) to correct bias arising from a model based on remote sensing satellite data on aerosol optical depth (AOD) and a deterministic chemical transport model, with this model generating annual estimates. 
Hammer et al. (2021) report this approach yielded cross-validated out-of-sample $R^2$ of 0.90-92 globally over a 21-year period of 2008-2018 \citep{MH2021, van_donkelaar_high-resolution_2015}.
The second model (GS2018; Figure \ref{fig:pm25_base}, Top Middle) fit a Bayesian hierarchical model to estimate annual average fine particle (PM2.5) concentrations at $0.1^{\circ} \times 0.1^{\circ}$ spatial resolution globally for 2010-2016 \citep{GS2018}. The model incorporated spatially varying relationships between 6003 ground measurements from 117 countries, satellite-based estimates, and other predictors. 
The authors quote a within-sample $R^2$ of 0.91 for the predictions from this model. The third model (SK2020; Figure \ref{fig:pm25_base}, Top Right) was a national-scale model for annual-average concentrations of PM$_{2.5}$ and several other pollutants from 1079-2015. This model applied partial least squares (PLS) to approximately 350 geographic characteristics at each monitoring site, including measures of traffic, land use, land cover, and satellite-based estimates of air pollution. The models combined the summary factors from this PLS representation with universal kriging to yield final estimates of annual PM$_{2.5}$ levels \citep{SK2020}. 
Taken together, these models incorporate different sources and forms of information on ambient PM$_{2.5}$ levels and are applied at different spatial scales. 

We apply \gls{BNE} to integrate these three models for PM$_{2.5}$ concentrations for 2011 in eastern New England, USA. In this application, \gls{BNE} generated \PM\ predictions that outperform those of existing approaches in terms of out-of-sample predictive accuracy. It also produced \textit{calibrated} estimates of predictive uncertainty, producing predictive intervals that have much lower \gls{NLL} and \gls{ECE} than the competing methods. Furthermore, we show that by leveraging the posterior estimate of \gls{BNE}'s different model components, we can decompose the model's overall predictive uncertainty into different sub-components 
(i.e., those due to ensemble weight estimation and those due to prediction) 
to understand the factors driving overall model uncertainty. 

%% file: sections/model.tex
\section{Model}
\label{sec:model}

\glsfirst{BNE} generalizes a classic ensemble model by augmenting the model's parametric components with Bayesian nonparametric machinery. Given observations $\{y_i, \bx_i\}_{i=1}^n$ for a response $y_i$ (\PM\ concentration) and feature space $\bx_i$ (spatial location: longitude and latitude), where $y \sim F^*(y|\bx)$ and a set of deterministic model predictions $\{f_k\}_{k=1}^K$, a classic ensemble model assumes the form:
\begin{align}
Y &= \sum_{k=1}^K f_k(\bx)\omega_k + \epsilon,
\label{eq:model_orig}
\end{align}
where $\bomega=\{\omega_k\}_{k=1}^K$ are the ensemble weights assigned to each base model, and $\epsilon$ is a zero-mean random variable describing the distribution of the outcome noise, which is commonly assumed to be Gaussian $\bepsilon \sim N(0, \sigma^2)$. 

The prediction of this classic Bayesian ensemble model is assessed using its  posterior predictive distribution. Denoting  $\Phi_\sigma(y|\bx, \mu)$ the \gls{CDF} of a Gaussian distribution, the posterior predictive distribution of a classic ensemble model is obtained by integrating $\Phi_\sigma(y|\bx, \mu)$ over the posterior distribution of the model parameter $\mu$:
\begin{alignat}{5}
F_{\mu, \sigma}(y|\bx) 
= \int_\real \Phi_\sigma(y|\bx, \mu)\pi_n(\mu, \sigma) d\mu d\sigma
\quad \mbox{where} \quad
\mu = \sum_{k=1}^K f_k(\bx)\omega_k,
\label{eq:pred_cdf}
\end{alignat}
where $\pi_n(\mu, \sigma)$ denotes the posterior distribution $(\mu, \sigma)$ given $\{y_i, \bx_i\}_{i=1}^n$, and we have denoted the predictive distribution function as $F_{\mu, \sigma}$ to stress its dependency on both the specification of the mean function $\mu=\sum_k f_k(\bx) \omega_k$ and of the variance $\sigma$. Then, the model's point prediction is expressed using the posterior mean $E(y|\bx)=\int y \, dF_{\mu, \sigma}(y|\bx)$, and the model's predictive uncertainty is expressed using the $(1-q)\%$ level credible interval $[F_{\mu, \sigma}^{-1}(\frac{q}{2}|\bx), F_{\mu, \sigma}^{-1}(1-\frac{q}{2}|\bx)]$.
Clearly, the quality of the predictive distribution $F_{\mu, \sigma}(y|\bx)$ depends both on the specification of mean function $\mu$ and the appropriateness of the distribution assumption in $\Phi_\sigma$. 
As a result, model biases in $\mu$ and $\Phi_\sigma$ negatively impact not only an ensemble model's predictive accuracy, but also its reliability in the quantifying the model's predictive uncertainty for the outcome $y$.

To achieve spatial adaptivity and mitigate model biases, \gls{BNE} augments both the mean and the predictive distribution function of a classic ensemble model using Bayesian nonparametric machinery. For the mean function $\mu$, \gls{BNE} replaces the constant weights $\{\bomega_k\}_{k=1}^K$ with random measures $\{\bomega_k(\bx)\}_{k=1}^K$. Therefore, it combines the base predictors  differently depending on the location of the input $\bx$ in the feature space, and then adds to $\mu$ a flexible residual process $\delta(\bx)$ to mitigate the systematic prediction bias shared among all the base predictors $f_k$'s. As a result, the mean function becomes:
\begin{alignat*}{5}
\mu = \sum_{k=1}^K f_k(\bx)\omega_k(\bx) + \bdelta(\bx).
\end{alignat*}
Then, to mitigate model biases in quantifying predictive uncertainty, \gls{BNE} models its predictive distribution function as a nonparametric random function $G$ centered around the original $F_{\mu, \sigma}$. So the final form of the \gls{BNE} is:
\begin{alignat}{5}
y|\bx \sim G \big[F_{\mu, \sigma}(y|\bx) \big]
\quad \mbox{where} \quad
\mu = \sum_{k=1}^K f_k(\bx)\omega_k(\bx) + \bdelta(\bx).
\label{eq:model_bne}
\end{alignat}
The random function $G$ can be considered as a flexible ``calibration function" that transforms the original parametric predictive \gls{CDF} $F_{\mu, \sigma}(y|\bx)$ toward the true data-generating \gls{CDF} $F^*(y|\bx)$ underlying the observed data, thereby relaxing the constant-variance Gaussian assumption encoded in $\Phi_\sigma$. 
The model parameters $\bomega$ (ensemble weights), $\bdelta$ (residual process) and $G$ (calibration function) are random functions following suitable Bayesian nonparametric priors.
In the following, Section \ref{sec:prior} introduces the prior specifications for the main model parameters $(\bomega, \bdelta, G)$. Section \ref{sec:hyperparameter} discusses the choices for the prior distributions and model hyperparameters. Finally, Section \ref{sec:model_prediction} derives the expressions for the \gls{BNE}'s predictive mean and variances in terms of model parameters, thereby making it explicit the roles that each model parameter plays in model prediction and uncertainty quantification. 
\subsection{Model Specification}
\label{sec:prior}

\subsubsection{Mean Function: Spatially Adaptive Ensemble}
\label{sec:sys_comp}

Recall that to achieve spatial adaptivity, BNE's mean function $\mu = \sum_{k=1}^K f_k(\bx)\omega_k(\bx)  + \delta(\bx)$ 
consists of the spatially adaptive ensemble weight $\bomega$, and a flexible residual process $\bdelta$ to mitigate the spatially varying systematic bias. The priors for $\bomega$ and $\bdelta$ are specified as:
\begin{align}
\bomega \sim & \; \mbox{Tailfree} (W, k_\omega), \quad
\bdelta \sim GP(0, k_\delta), 
\label{eq:prior_sys}
\end{align}
respectively. Specifically, $\bomega_k(\bx)$ is a collection of random measures that controls the contribution of each individual base model $f_k$ to the overall ensemble. We model $\bomega$ as the multivariate logistic transformation of $K$ independent Gaussian processes corresponding to each $f_k$:
\begin{align}
\bomega_k(\bx) &= 
\frac{exp \big( w_{k}(\bx) \big)}
{ \sum_{k'=1}^{K} exp \big( w_{k'}(\bx) \big)}, 
\qquad
\{ w_k \}_{k=1}^K \stackrel{iid}{\sim} GP(\bzero, k_{\omega}).
\label{eq:prior_omega}
\end{align}
Denoting the collection of Gaussian processes $\{ w_k \}_{k=1}^K$ as $W$, we have specified a dependent tail-free process (DTFP) prior \citep{jara_class_2011} for the ensemble weights, as specified in Equation~\ref{eq:prior_sys}.

We model $\delta$ nonparametrically using a \gls{GP} with zero mean function $\bzero(x)=0$ and kernel function $k_\delta(\bx, \bx')$. As a result, in densely-sampled regions that are well captured by the training data, $\delta(\bx)$ will confidently (i.e., with low posterior variance) mitigate the prediction bias between the observation $y$ and the prediction function $\sum_{k=1}^K f_k(\bx)\omega_k(\bx)$. However, in sparsely-sampled regions, the posterior mean of $\delta(\bx)$ will fall back to $\bzero(x)=0$ so as to leave the predictions of the original ensemble intact (since these expert-built base models presumably have been specially designed for the problem being considered), and the posterior uncertainty of $\delta(\bx)$ will increase to reflect the model's increased uncertainty at location $\bx$ that is far away from the observations. 

\subsection{\gls{HEX-GP}}
\label{sec:link_func}

To flexibly capture the complex spatial variation in the distribution of ambient air pollution, BNE models its predictive distribution $F(y|\bx)$ as a flexible transformation of the original predictive distribution function $F_{\mu, \sigma}$:
\begin{align}
F(y|\bx) =  G \big[ F_{\mu, \sigma}(y|\bx) \big],
\label{eq:calib_func}
\end{align}
where $G$ is a spatially-adaptive "calibration function" following a suitable Bayesian nonparametric prior. To design a proper $G$ model, we notice that in air pollution modeling, data observations (e.g., measurements from air pollution monitors) are distributed sparsely and unevenly across geographical regions, yet the total sample size can be large due to the number of total monitors and, potentially, repeated measurements. As a result, preference is placed on parsimony that offers both computationally tractability and sufficient flexibility to capture the spatially varying patterns of dispersion and skewness.

To this end, we present a tractable extension to the classic Gaussian process, \gls{HEX-GP}. \gls{HEX-GP} augments a Gaussian process model with flexible Bayesian nonparametric estimators for its higher moments, and deploys smooth shrinkage on these moment estimators to guard against overfitting. Specifically, given a standard Gaussian process model
\begin{align*}
    y | \bx \sim N(\mu(\bx), \; \sigma^2), \quad \mu(\bx) \sim GP(f_0, k_{\mu}),
\end{align*}
where $f_0$ is the prior mean function of the Gaussian process.

\gls{HEX-GP} models the spatially-heterogeneous variance in $y|\bx$ by introducing a non-negative scaling factor $s(\bx)$ to its variance component, such that:
\begin{align*}
    y | \bx \sim N(\mu(\bx), \; \sigma^2 \underline{* s^2(\bx)}),
\end{align*}
where $s(\bx)$ follows a Bayesian non-parametric prior that is \textit{a priori} centered around 1. Then, to model the potential patterns of skewness in data distribution, we introduce an \textit{exponential-modification} $m(\bx)$ to the mean function:
\begin{align}
    y | \bx \sim N(\mu(\bx) \underline{ + m(\bx)},\; \sigma^2 \underline{* s^2(\bx)}); \quad
    m(\bx) \sim Exp(\lambda(\bx)),
    \label{eq:hex_gp_skew}
\end{align}
where $\lambda(\bx)$ is the scale parameter of the exponential distribution. Notice that if holding $(\mu, \sigma, s)$ constant, (\ref{eq:hex_gp_skew}) leads to the well-known \gls{EMG} distribution for skewed continuous data on the real support $\real$ \citep{Grushka1972CharacterizationOE}. The \gls{EMG} distribution is able to exhibit a Pareto-like (fat) right tail, and reverts back to a Gaussian shape as $\lambda \rightarrow 0$ \citep{sager2019double}.

Under (\ref{eq:hex_gp_skew}), we specify the following log priors for the random functions $(s, \lambda)$:
\begin{align}
    log \, \lambda \sim GP(\mu_\lambda, \sigma^2_\lambda * k_\lambda); \quad log \, s \sim GP(\textbf{0}, \sigma^2_s * k_s),
    \label{eq:hex_gp_prior}
\end{align}
where $\textbf{0}$ is the constant zero function, and $\mu_\lambda(\bx)$ is a constant function with large negative value, so that  $\lambda(\bx)$ is centered \textit{a priori} to $\varDelta$ which is a small value close to 0. Here $(\sigma^2_\lambda, \sigma^2_s)$ are pre-specified hyper-priors  that control the strength of model shrinkage of the nonparametric estimators $(m(\bx), s(\bx))$ towards their prior centers $\varDelta$ and 1.

Coming back to (\ref{eq:calib_func}), the \gls{HEX-GP} induces a calibration function $G$ that takes a tractable form:
\begin{align*}
    G \big[ F_{\mu}(y|\bx) \big] = 
    F_{\mu}(y|\bx) - 
    C(\bx) * 
    e^{-\frac{y - \mu(\bx)}{\lambda(\bx)}} * 
    F_{\mu} \Big( y - \frac{\sigma^2 s^2(\bx)}{\lambda(\bx)} \big|\bx \Big),
\end{align*}
where $C(\bx) = exp \big( \frac{\sigma^2 s^2(\bx)}{2\lambda^2(\bx)} \big)$ is a positive constant. As constructed, $G$ applies a sophisticated additive modification to the original distribution function $F_\mu$, shifting the probability mass in a positive direction by reducing the value of $F_\mu$ when $y$ is small. The degree of modification is modulated by the magnitude of $\lambda$, and disappears as $\lambda \rightarrow 0$.

\subsection{Kernel Functions and Hyperparameters}
\label{sec:hyperparameter}
Throughout this work, we model $k_{\omega}$, $k_{\delta}$ and $(k_{s}, k_{\lambda})$ using the Mat\'{e}rn $\frac{3}{2}$ kernel family with a length-scale parameter $l \geq 0$:
\begin{align}
    k(\bz, \bz'|l) = 
    \Big(1 + \frac{\sqrt{3}||\bz - \bz'||_2}{l} \Big) 
    * 
    exp \Big( -\frac{||\bz - \bz'||_2}{l} \Big).
    \label{eq:rbf}
\end{align}
We denote the length-scale parameters for $k_{\omega}$, $k_{\delta}$ and $k_{G}$ as $l_\omega$,  $l_\delta$, and $l_G$, respectively.

Mat\'{e}rn $\frac{3}{2}$ kernels are well suited for quantifying the predictive uncertainty of a given model, since they produce predictive variances that are explicitly characterized by the distance of a given point from the training data. Therefore, for a model with a Mat\'{e}rn $\frac{3}{2}$ kernel, its predictive variances increases as the prediction location moves further away from the training data. 
Furthermore, a Mat\'{e}rn $\frac{3}{2}$ kernel comes with a theoretical guarantee in estimating distribution parameters. 
That is, the sample space of a Mat\'{e}rn $\frac{3}{2}$ process corresponds to the space of H\"{o}lder continous functions that are at least once differentiable, allowing $(\delta, s, \lambda)$ to flexibly capture the spatially adaptive biases and overdispersions in the data \citep{van_der_vaart_reproducing_2008, van_der_vaart_information_2011}.

The hyperparameters for a \gls{BNE} are $\sigma^2$ (the noise variance)
the kernel hyperparameters $\{l_\omega, l_\delta, l_s, l_\lambda\}$, 
and the distribution-parameter regularizers $(\mu_\lambda, \sigma_\lambda, \sigma_s)$.
In practice, to guarantee a proper rate of posterior convergence, it is important that the model hyperparameters are estimated adaptively \citep{van_der_vaart_adaptive_2009}. 
Consistent with existing \gls{GP} approaches \citep{team_stan_2018}, we place the inverse-Gamma priors on the $l$'s and the Half Normal priors on $\sigma$, i.e.:
\begin{align*}
l_\theta & \sim invGamma(\alpha_\theta, \beta_\theta) \quad \mbox{for} \; \theta \in \{\omega, \delta, s, \lambda\},
\end{align*}
where $\alpha$, $\beta$ where chosen so the prior probability for $l$ falling into a desired range $[l_{lower}, l_{upper}]$ is high. In this work, we set $l_{lower}=2$ and $l_{upper}=10$ such that $P_{invGamma}(l \in [2, 10]|\alpha, \beta)=0.98$ for all length-scale parameters. For the variance parameter $\sigma$, we use the weakly informative half-Gaussian prior:
\(
\sigma \sim HalfNormal(0, 5).
\)
Finally, to regularize the variance and skewness parameters $(s, \lambda)$, we set $\mu_\lambda=-3$ and 
\begin{align*}
\sigma_s \sim HalfNormal(0, 1), \quad \sigma_\lambda \sim HalfNormal(0, 1),
\end{align*}
thereby encouraging shrinkage on the random functions $(s, \lambda)$ by putting nontrivial prior mass around zero.


\section{Prediction and Uncertainty Quantification}
\label{sec:model_prediction}

The model parameters for \gls{BNE} are $\bomega$ (the ensemble weights), $\bdelta$ (the residual process) and $G=(s, \lambda)$ (the calibration function). Denoting $\bar{\beta}$ the posterior mean  of a BNE model parameter $\beta \in \{\bomega, \delta, s, \lambda \}$, the predictive mean of the \gls{BNE} is:
\begin{align}
E(y|\bx) &= 
\sum_{k=1}^K f_k(\bx) \bar{\omega}_k(\bx) + 
\bar{\delta}(\bx)
+ 
\int_{y \in \Ysc} 
\Big[ F_\mu(y|\bx) - \bar{G} \big[ F_\mu(y|\bx) \big] \Big] dy  \nonumber
\\
&= \sum_{k=1}^K f_k(\bx) \bar{\omega}_k(\bx) + \bar{\delta}(\bx) + \bar{\lambda}(\bx)
\label{eq:bne_pred_mean}
\end{align}
This results shows that the predictive mean for a \gls{BNE} is composed of three components: (1) the predictive mean of the original adaptive ensemble $\sum_{k=1}^K f_k(\bx)\bar{\omega}(f_k, \bx)$; (2) the term $\bar{\bdelta}$ representing \gls{BNE}'s ``direct" correction to the prediction function obtained through estimation of the residual process $\bdelta$, and (3) the term $\int \big[ F_\mu(y|\bx) - \bar{G} [ F_\mu(y|\bx) ] \big] dy=\bar{\lambda}(\bx)$ representing a \gls{BNE}'s ``indirect" correction to the prediction function obtained upon relaxation of the Gaussian assumption. 


To understand the behavior of BNE's model uncertainty, we decompose $Var(y|\bx)$ using the law of total variance:
\begin{align}
    Var(y|\bx) &= 
    \underbrace{
    E \big[ Var(y|\bx, (\bomega, \bdelta, G)) \big]}_{aleatoric \;\; uncertainty} 
    + 
    \underbrace{
    Var \big[ E(y|\bx, (\bomega, \bdelta, G)) \big]}_{epistemic \;\; uncertainty}  
    \nonumber \\
    &= 
    E\Big[
    \sigma^2 \, s^2(\bx) + \lambda^2(\bx) 
    \Big] + 
    Var \Big[ 
    \sum_{k=1}^K f_k(\bx) \omega_k(\bx) + \delta(\bx) + \lambda(\bx)
    \Big].
    \label{eq:bne_var}
\end{align}
Given a location $\bx$, the first component $E\Big[\sigma^2 \, s^2(\bx) + \lambda^2(\bx) \Big]$ describes the model's posterior estimate of the \textit{aleatoric uncertainty} (i.e., the observed stochasticity in the data distribution), whereas the second component $Var \big[ \sum_{k=1}^K f_k(\bx) \omega_k(\bx) + \delta(\bx) + \lambda(\bx) \big]$ is the model's posterior variance reflecting the \textit{epistemic uncertainty}, i.e., the uncertainty due to the lack of similar examples in the training data \citep{der2009aleatory, sullivan_introduction_2015, hullermeier2021aleatoric}. As a result, in the regions containing  training data, the model flexibly captures the heterogeneous patterns in the data distribution while maintaining a low level of epistemic uncertainty. On the other hand, in regions outside the training domain, the aleatoric uncertainty reverts back to $\sigma^2$ as the adaptive components reverting to their prior $(s, \lambda) \rightarrow (1, \varDelta)$, while the epistemic uncertainty increases as $\delta$ reverts back to their Gaussian process priors with high variance. 

From the perspective of uncertainty quantification, each component of (\ref{eq:bne_var}) plays a crucial role: without the aleatoric component, the model loses the ability to capture the spatially adaptive patterns in the data due to variance (via $s^2(\bx)$) or heavy-tailness (via $\lambda^2(\bx)$). In this case, the model-based confidence interval will only reflect the model's epistemic uncertainty. This can be problematic in well-observed regions with above-average levels of noise in the observed data. In this case,  the confidence intervals will fail to achieve nominal coverage due to low-degree of epistemic uncertainty. On the other hand, without the epistemic component (esp. the posterior variance in $\delta(\bx)$), the model loses the ability to reflect large amounts of uncertainty in regions having little training data, which may lead to severely over-confident predictive intervals that are too narrow to cover the observed data.

%% file: sections/inference_v1_nofig.tex
\section{Model Fitting and Inference}
\label{sec:inference}

For a Bayesian model that specifies a model space for the predictive distribution  $\Msc=\{F(y|\bx)\}$, model estimation finds the projection of the empirical distribution $\Fbb(y|\bx)$ of the data in the model space $\Msc$. To this end, a \gls{BNE} specifies a large model space that is centered around a naive ensemble model $Y = \sum_{k=1}^K f_k(\bx)\omega_k(\bx) + \epsilon$. Specifically, as shown in Figure \ref{fig:model_space}, an \gls{BNE} starts from the naive model space assuming no prediction bias and a Gaussian outcome:
\begin{align*}
\Msc_{\omega}=
\Big\{F(y|\bx) \Big|F=F_\mu(y|\bx), \mu=\sum_{k=1}^K f_k(\bx)\omega_k(\bx) \Big\}.
\end{align*}
Then, by adding $\delta$, an \gls{BNE} expands $\Msc_{\omega}$ to a model space $\Msc_{\omega, \delta}$ using the bias-corrected (i.e. nonparametric) prediction function $\mu$, still under the Gaussian assumption for the outcome:
\begin{align*}
\Msc_{\omega, \delta}=
\Big\{F(y|\bx) \Big|
F=F_\mu(y|\bx), 
\mu=\sum_{k=1}^K f_k(\bx)\omega_k(\bx) + \delta(\bx) \Big\}.
\end{align*}
Finally, using $G$, an \gls{BNE} expands to the full model space $\Msc_{\omega, \delta, G}$, which represents a nonparametric prediction function based on minimal distribution assumptions:
\begin{align*}
\Msc_{\omega, \delta, G}=
\Big\{F(y|\bx) \Big|
F=G\big[F_\mu(y|\bx) \big], 
\mu=\sum_{k=1}^K f_k(\bx)\omega_k(\bx) + \delta(\bx) \Big\}.
\end{align*}
It is easy to see that the three model spaces are nested, i.e. $\Msc_{\omega} \subset \Msc_{\omega, \delta} \subset \Msc_{\omega, \delta, G}$. 

Notice that for nonparametric estimation, naively projecting $\Fbb(y|\bx)$ to the full model space $\Msc_{\omega, \delta, G}$ (i.e. fitting the full \gls{BNE} model directly) can lead to non-identifiability of the model parameters. In particular, since $\bomega$ and $\bdelta$ are non-identifiable in the mean component $\mu$, estimation of the full \gls{BNE} model directly can result in the undesirable situation where the flexible residual process $\bdelta$ overfits the training data and "squeezes" the estimates for the ensemble weight $\bomega$ to a suboptimal value. This not only negatively impacts the model's prediction performance, it also makes it difficult to distinguish between the discrepancy of the base models and the effects of the nonparametric calibration parameters \citep{arendt_quantification_2012, brynjarsdottir2014learning, tuo_efficient_2015}.

Consequently, to ensure the identifiability, in this work we propose a step-wise algorithm that performs model estimation by finding projections in a sequentially expanding model space $\Msc_{\omega} \subset \Msc_{\omega, \delta} \subset \Msc_{\omega, \delta, G}$, thereby placing implicit regularization on model parameters in each step.  We show that the proposed algorithm is computationally convenient in that it  decomposes the estimation of the full \gls{BNE} into three classic sub-problems 
(Section \ref{sec:gibbs}). 

\subsection{The Algorithm}
\label{sec:gibbs}
Briefly, the algorithm first initializes $\delta$ and $G$ to their default values $\delta(\bx)=0$ (the zero function) and $G(\bz)=\bI(\bz)$ (the identity function, achieved by setting $(\lambda, s) = (0, 1)$), then updates the model parameters sequentially. 
That is, it updates $\bomega$ holding $\delta=\bzero, G=\bI$, then updates $\bdelta | \bomega$ holding $G=\bI$, and so on. The approach first finds the ensemble weights $\bomega$ that best explain the data with a model assuming no prediction bias (i.e. $\delta=\bzero$) and under the Gaussian assumption for the outcome (i.e. $G=\bI$), then estimates $\bdelta(\bx)$ to correct for residual bias under the Gaussian assumption, and finally estimates $G$ to relax the Gaussian assumption for the outcome. The algorithm concludes after estimating $G$. 
Specifically:

\begin{enumerate}
    \item[\textbf{Step 1}:] \textbf{Adaptive Ensemble}: \textbf{Update} $\bomega$
    
    This step fits the naive ensemble model $Y = \sum_{k=1}^K f_k(\bx)\omega_k(\bx) + \epsilon$, as specified by $\Msc_\omega$. The model likelihood is:
    \begin{align*}
        f(\omega | \{y_i, \bx_i\}) \propto 
        exp\bigg(
        -\frac{1}{2 \sigma^2} 
        \Big(y_i - \sum_{i=1}^k f_k(\bx_i)\omega(\hat{f}_k, \bx)\Big)^2
        \bigg)
        \prod_{k=1}^K f(w_k),
    \end{align*}
    where $f(w_k)$ are the Gaussian process likelihoods for the DTFP prior of $\bomega$ in (\ref{eq:prior_omega}).
    
    Since $\{f_k(\bx_i)\}_{k=1}^K$ are deterministic, this step can be understood as a linear regression but with the regression coefficients $\bomega$ adaptive to the input $\bx_i$ \citep{pati_posterior_2013}.
    
    \item[\textbf{Step 2}:] \textbf{Bias Correction}: \textbf{Update} $\bdelta | \bomega$
    
    This step expands the model space from  $\Msc_\omega$ to $\Msc_{\omega, \delta}$ by fitting the bias-corrected ensemble model conditional on $\bomega$:
    \begin{align}
        Y = \sum_{k=1}^K f_k(\bx)\omega(f_k, \bx) + \bdelta(\bx) + \epsilon.
        \label{eq:gibbs_step2_model}
    \end{align}
    
    This step can be understood as running a Gaussian process regression. 
    This is because conditional on $\omega$, (\ref{eq:gibbs_step2_model}) is a Gaussian process model with mean function $\mu_\omega = \sum_{k=1}^K f_k(\bx)\omega(f_k, \bx)$ and kernel function $k_\delta$. 
    The posterior distribution for $\delta|\bomega$ is a Gaussian process:
    \begin{align*}
        \bdelta | \bomega 
        &\sim GP \Big(\mu_{\bdelta | \bomega},  \bK_{\bdelta | \bomega}\Big), 
        \qquad \mbox{where} 
        \\
        \mu_{\bdelta | \bomega} &= 
        \mu_\omega + \bK_\delta(\bK_\delta + \sigma^2\bI)^{-1}(\by - \mu_\omega) \qquad \mbox{and} 
        ,
        \\
        \bK_{\bdelta | \bomega} &= 
        \bK_\delta - \bK_\delta(\bK_\delta + \sigma^2 \bI)^{-1}\bK_\delta,
    \end{align*}
    where $\bK_\delta$ is the kernel matrix constructed using kernel function $k_\delta$.
    
    \item[\textbf{Step 3}:] \textbf{Distribution Calibration}: \textbf{Update} $G, \bdelta | \bomega$ 

    This step expands $\Msc_{\omega, \delta}$ to the full model space $\Msc_{\omega, \delta, G}$ by estimating the calibrated predictive  distribution function $F(y|\bx) = G[F_\mu(y|\bx)]$, where $F_\mu(y|\bx)$ is the predictive distribution function generated by the bias-corrected ensemble model under Gaussian distribution assumption from Step 2. 
    
    To estimate $G=(s, \lambda)$ and update $\delta$, we initialize from $\Msc_{\omega, \delta}$, and run a standard inference algorithm with respect to the hierarchical model:
    \begin{align}
    \label{eq:bne_full_likehood}
    y|\bx &\sim N(\bmu_\omega(\bx) + \delta(\bx) + m(\bx), \sigma^2 * s^2(\bx)) \\
    \delta &\sim GP(0, k_\delta), \quad log \, s \sim GP(0, \sigma_s^2 * k_s ) 
    \nonumber \\
    m &\sim Exp(\lambda), \quad log \, \lambda \sim GP(\mu_\lambda, \sigma^2_\lambda * k_\lambda),
    \nonumber 
    \end{align}
    where $(\bmu_\omega, \sigma^2)$ follows the posterior distribution from $\Msc_{\omega}$, and $(\sigma_s^2, \sigma_\lambda^2)$ are the pre-specified hyper-parameters controlling the strength of model skrinkage onto $(\lambda, s)$. Notice that under BNE, the mean-parameter convolution $\delta(\bx) + m(\bx)$ do not lead to non-identifiability issues. This is because the $\lambda$ parameter is controlled by the skrinkage prior (\ref{eq:hex_gp_prior}) and can be uniquely identified in the presence of skewness and heaviness of the tail of the data distribution.
\end{enumerate}

A comment regarding computation is in order. As shown, the proposed algorithm decomposes the full estimation problem into three sub-problems, where the model progressively imposes layers of Bayesian nonparametric augmentations to address potential misspecification while maintaining identifiability.
The posterior distributions for each of these sub-problems either have closed-form, or are smooth and differentiable functions of model parameters that can be sampled efficiently using standard gradient-based MCMC methods such as the \gls{NUTS} \citep{hoffman_no-u-turn_2011}. However, for larger data sets, the issue of computational scalability needs to be addressed. 
For example, the sample size of a spatio-temporal dataset is typically many times larger than that of a spatial dataset, which presents significant computational challenges to a \gls{GP} due to its computational complexity of $O(n^3)$. 
To this end, in Appendix \ref{sec:rff} we supply a 
practical random-feature method that approximates the \gls{GP} posterior with $O(n)$ computational complexity \citep{rahimi_random_2008}. This approach, combined with minibatch-based stochastic gradient MCMC algorithms, can scale to millions of observations \citep{JSSv091i03}.



%% file: sections/simu_nofig.tex
\section{Simulation Studies}
\label{sec:exp}

We investigate the performance of the proposed method on a collection of simulated time-series and spatial regression datasets. 
We measure the model's predictive accuracy using the \gls{MSE}, and we measure the model quality in uncertainty calibration with two metrics: the \glsfirst{NLL} of the model's posterior predictive distribution, and also the  \glsfirst{ECE}, which measures the empirical coverage of the model's predictive intervals. 

Specifically, to compute \gls{NLL}, we notice that all models in the experiment can be written in a form that adopts an Gaussian outcome. As a result, given $p$, the joint posterior distribution of $(\mu, \sigma)$, the posterior predictive negative log likelihood is:
\begin{align}
    NLL &= \int_{\mu, \sigma} \Big[ 
    \underbrace{\frac{(y - \mu(x))^2}{\sigma^2}}_{calibration} + 
    \underbrace{log \sigma \vphantom{\frac{a}{c}}}_{sharpness}
    \Big] dp(\mu, \sigma).
    \label{eq:nll}
\end{align}
NLL simultaneously measures the model prediction's quality in \textit{calibration} 
and in \textit{sharpness} 
\citep{gneiting_probabilistic_2007}. \gls{NLL} belongs to the family of \textit{proper scoring rules}, a well-established family of metrics for assessing a model's calibration quality in the probabilistic forecast literature 
\citep{10.2307/2984087, bernardo1979expected}.

On the other hand, given a model with predictive CDF $F(y|\bx)$, we compute \gls{ECE} by first computing the model's $q\%$ predictive intervals $CI_q=\big[ F^{-1}(\frac{q}{2}|\bx_i), F^{-1}(1-\frac{q}{2}|\bx_i) \big]$ for all observations $\{y_i, \bx_i\}_{i=1}^N$, and estimate the empirical coverage probability as $ \hat{q}(F) = \frac{1}{N} \sum_{i=1}^N \bbI \big(y_i \in CI_q(\bx_i | F) \big)$. If the model's predictive distribution properly captures the empirical distribution of the data, the empirical coverage probability $\Pbb(y \in CI_q | F)$ will be close to the nominal coverage probability $q$, regardless of the value of $q$. To this end, \gls{ECE} summarizes the predictive model's quality in uncertainty quantification using the squared difference between $q$ and $\Pbb(y \in CI_q | F)$ for $q \in [0, 1)$, 
\begin{align}
    ECE &=  \int_{0}^1 || q - \Pbb(y \in CI_q | F) ||_2^2 \; dq
    \label{eq:ece}
\end{align}

We compare the performance of \gls{BNE} with four other commonly used ensemble algorithms, summarized in Table \ref{tab:exp_methods}. These are 
(1) the Bayesian stacking (\textbf{stack}), which aggregates models using non-adaptive weights but calibrates the predictive distribution using leave-one-out cross validation \citep{yao_using_2018}; 
(2) the generalized additive model ensemble (\textbf{GAM}), which combines the base-model
predictions in a regression model that includes additive, linear terms for each set of predictions and an extra smoothing spline term as a function of the feature space to mitigate any systematic bias \citep{xiao_ensemble_2018};
(3) adaptive Bayesian model averaging (\textbf{BMA}), which  aggregates models using spatially adaptive weights parameterized using Gaussian processes; and (4) Bayesian adaptive ensemble (\textbf{BAE}), which mitigates the predictive bias in BMA using an additive Gaussian process \citep{murray_bayesian_2019}. The BMA and BAE are reduced versions of \gls{BNE}, where BMA corresponds to a \gls{BNE} model with no bias correction and no distribution calibration and BAE corresponds to a \gls{BNE} model with no distribution calibration.

We compare the performance of these competing models when applied to both time-series and spatial data having complex mean structure and heterogeneous variability. 
For the time-series data, we consider a nonlinear time series 
$f=sin(x)$,  with $x \in (-5\pi, 5\pi)$. We consider a data distribution of $y \sim logNormal(f(x), \sigma(x))$ with  $\sigma(x)=exp(cos(x) - 1)$ so that it exhibits heterogeneous variance and skewness depending on the location of the input space (see Figure \ref{fig:base_exp_1d}). 
For a spatial scenario, we use the multimodal Mishra's Bird function $f(x_1, x_2) = cos(4 x_1) * exp\big( (1 - sin(4 x_2))^2 \big) + sin(4 x_2) * exp\big( (1 - cos(4 x_1))^2 \big) + (x_1 - x_2)^2$ for $x_1, x_2 \in (-\pi, \pi)$. This data generating function, common in the machine learning and optimization literature, is characterized by a slow-varying global trend but with sharp local fluctuations  \citep{mishra_new_2006} (Figure \ref{fig:base_exp_2d}). 
See Section \ref{sec:exp_spp_data_gen} (Supplementary Material) for further detail.

Tables \ref{tb:exp_1d_res} and \ref{tb:exp_2d_res} reports the operating characteristics (\gls{MSE}, \gls{NLL}, \gls{ECE}
, and coverage probability of $95\%$ \gls{CI}
) for each ensemble method in the time-series and spatial settings, respectively.
An important goal of this simulation study is to understand if the quality of the ensemble model's predictive uncertainty remains robust even in regions that are outside training data. To this end, we test the model's uncertainty performance in an input space that is larger than that covered by the training data. That is, a test time window of $(-5\pi, 5\pi)$ versus a training time window of $(-3\pi, 3\pi)$ in the time-series experiments, and a test region of $(-1.25\pi, 1.25\pi)^2$ versus a training region of $(-\pi, \pi)^2$ in the spatial experiments. Ideally, we expect the ensemble model to provide not only accurate prediction for test samples from the training region (as measured by \gls{MSE}), but also calibrated uncertainty quantification across the whole test region even if it is not covered by the training data (as measured by \gls{NLL}, \gls{ECE} 
, and coverage probability of $95\%$ \gls{CI}s
).

As shown, the relative performances of the various ensemble models are largely consistent across both scenarios. Specifically, as the sample size increases, both the \gls{MSE} and \gls{ECE} for \textbf{stack} and \textbf{BMA} are higher than that of the other approaches, due to their inability to correct for bias in the base models.   \textbf{GAM} and \textbf{BAE} improved upon the previous two methods by adapting the mean prediction to the data. However, the inflexibility 
in the distributional assumptions made by these approaches prevented them from capturing the heterogeneity and skewness in the error distribution, producing sub-optimal uncertainty estimates (both in terms of the proper scoring rule, as measured by \gls{NLL}, and coverage of predictive interval, as measure by \gls{ECE}) even in large samples. On the other hand, \textbf{BNE} improved with increasing sample size due to the flexibility in its distributional assumptions. 

To further understand the behavior of these models, we decompose the \gls{NLL} score into its calibration and sharpness components (i.e., $\frac{(y-\mu(x))^2}{\sigma(x)^2}$ and $log \sigma$, respectively), and also measure the coverage probability of the models $95\%$ predictive credible intervals, both in the training region and in the test region. Tables \ref{tb:exp_1d_res_app}-\ref{tb:exp_2d_res_app} in the Supplementary Material shows the results. As shown, compared to models with inflexible distribution assumptions, the \textbf{BNE} model achieves both better calibration and on average narrower credible intervals  (i.e., better sharpness). At the same time, the \textbf{BNE}'s $95\%$ credible intervals better maintain nominal coverage, both in the training region and across the entire test region. This advantage is especially prominent in the case of small samples ($n=10$) and in the case of complex spatial heterogeneity (i.e., in the spatial experiments). This result shows that the \textbf{BNE} model is not simply  improving model uncertainty by widening its credible intervals everywhere, and the \textbf{BNE}'s flexible distributional model does not lead to over-fitting. Instead, this approach better captures complex patterns in the training data, and better expresses epistemic uncertainty outside of the training regions.

%% file: sections/application_nofig.tex
\section{Spatial ensemble of \PM\ levels in Eastern New England, USA}
\label{sec:application}

Here, we use \gls{BNE} to integrate information on \PM\ levels from the three distinct \PM\ exposure models introduced in Section \ref{sec:data}, 
MH2021, GS2018, and SK2020, and produce a single set of predicted concentrations for average \PM\ levels in 2011.   
Our goals are to (1) improve prediction accuracy relative to any one set of predictions; (2) understand the driving factors behind the ensemble system's uncertainty; and (3) understand spatial locations where the available exposure models systematically underperform.
We implement our ensemble framework described in Sections \ref{sec:model} and \ref{sec:inference} on the three exposure models' out-of-sample prediction for 51 monitors, along with the other four ensemble approaches that were also considered in the simulation studies. For all Bayesian methods, we perform posterior inference using \gls{HMC} with $10^4$ MCMC samples after using a burn-in of $1000$. 

In order to mitigate information leakage caused by autocorrelation among neighboring monitors, we employed spatial cross-validation techniques to assess performance of the various methods. Specifically, we applied the K-means algorithm to cluster all the available data. We divided the dataset consisting of 51 monitors into 6 distinct folds, with each fold containing 6, 10, 8, 9, 9, and 9 monitors respectively. During the grid search, we considered all three metrics, and the outcomes are presented in Table 4. Additionally, we visually represented the predictive uncertainty of BAE (posterior predictive variance) for the results obtained from the test set of one particular fold, as depicted in Figure \ref{fig:spcv}.

Table \ref{tb:app_comp} compares the overall predictive performance of the BNE model to that of existing methods.  The posterior estimates from the BNE model consistently returned the lowest \gls{RMSE}, 
indicating improved quality in both predictive accuracy and uncertainty calibration when compared to that of the existing approaches considered. The bottom left, center, and right panels present the posterior estimates of the spatially-adaptive ensemble weights for the MH2021, GS2018, and SK2020 models, respectively,  from the \gls{BNE} fit.
For this study region, results show that the predictions from the SK2020 model are weighted the most, except in the western-most edge of the region, where MH2021 is weighted most. In general, 
\gls{BNE} assigns low weights to the GS2018 model throughout the region, suggesting this model does not predict the held-out data as well as the other two exposure products.  One potential reason for this finding is it is apparent the GS2018 model is not as highly resolved spatially as the other two, which can be observed from the resolution depicted in the uppper panel in Figure \ref{fig:pm25_base}. 


Figure \ref{fig:app_res} visualizes the posterior predictive distribution for \PM\ levels over the study region from the \gls{BNE} fit. Specifically, Figures \ref{fig:app_post_mean}--\ref{fig:app_post_unc} show \gls{BNE}'s posterior predictive mean and predictive uncertainty (posterior predictive variance), respectively. Figures \ref{fig:app_post_uncn_ens}-\ref{fig:app_post_uncn_resid} show the decomposition of the predictive uncertainty into that due to model uncertainty (i.e. the posterior variance associated with the ensemble weights $\omega$) and that due to model prediction (i.e.  the posterior variance for $\delta$ and $G$), respectively. In all figures, the black circles indicate the locations of the air pollution monitors.  \ref{fig:app_post_uncn_ens} shows higher levels of uncertainty in the north and northwest of the study region, and also in regions that are farther away from this urban areas where monitoring data is sparse.   Comparing these uncertainty estimates to the model-specific predictions in Figure \ref{fig:pm25_base}, we notice that the north and northwest areas are those where the MH2021 and SK2020 models tend to disagree.  

%% file: sections/conclusion.tex
\section{Conclusion}
\label{sec:conclusion}

In this work, we have presented a Bayesian nonparametric approach for combining deterministic model predictions to construct a \textit{spatially adaptive} ensemble that produces  \textit{calibrated}  predictive uncertainty. 
To illustrate the practical effectiveness of the proposed approach, we compared via simulation the performance of the \gls{BNE} model with that of several existing ensemble methods. This comparison showed clear improvements in both out-of-sample predictive accuracy and in uncertainty quantification. Finally, we applied the proposed approach to integrate three air pollution prediction models for predicting annual \PM\ concentrations in eastern New England. The resulting predictions outperformed those of the existing ensemble approaches in terms of out-of-sample \gls{RMSE}.

By producing a set of exposure predictions with rigorously quantified uncertainty, the proposed \gls{BNE} model can be used to provide calibrated exposure estimates for downstream analyses.  For instance, there is currently great interest in identifying characteristics of locations (e.g. urban, rural, region of the country) and times (e.g. seasons or years) most associated with high uncertainty of of air pollution estimates, and how such varying prediction error may impact downstream epidemiologic health effects analyses.   A few recent efforts have sought to develop measurement-error corrections applicable to such spatio-temporally varying measurement error \citep{alexeeff_spatial_2016}, and the uncertainty estimates obtained from our proposed framework can serve as inputs into such approaches.  Moreoever, reporting
the estimated model weights and uncertainty estimates from this approach back to teams developing the original prediction models can inform efforts to improve these models. For instance, identification of high uncertainty areas can help inform future monitoring campaigns or suggest additional model inputs that could improve model performance. 

The application we considered focused on the integration of \textit{spatial} prediction models for annual \PM\ averages. 
In practice, the such integration of multiple \textit{spatio-temporal} air pollution prediction models
is also of interest in many applications.  Adapting \gls{BNE} to the spatio-temporal setting is conceptually straightforward. We can incorporate time as an additional input to the ensemble weight $\omega_k(\bx)$. However, this spatio-temporal  setting raises a few novel challenges 
that require careful treatment for the model to remain effective. The first is additional model complexity introduced by the temporal dimension. For spatio-temporal data, the ensemble model needs to be sufficiently flexible to handle  temporal trends in air pollution concentrations, such as seasonal variability, that are distinct from its spatial variability.  
In the context of \gls{BNE}, these two objectives can be achieved by designing a temporal kernel function $k_{time}(t, t')$ that models the temporal dimension using smoothing temporal basis functions similar to those used in classic spatio-temporal models for air pollution \citep{keller_joshua_p_unified_2015, lindstrom_flexible_2014, szpiro_predicting_2009}. Then, the spatio-temporal dependency can be modelled by an interaction kernel $k_{space \times time} = k_{space} \otimes k_{time}$ that is constructed as a tensor product, and is regularized by certain sparse-inducing priors (e.g., horseshoe), leading to a final kernel in the form of $k = k_{space} + k_{time} + k_{space \times time}$ \citep{gu_smoothing_2013, li_variable_2017}. We leave the exploration for the optimal kernel design for modeling spatio-temporal interactions to future work. 

The distribution model developed in this work (\gls{HEX-GP}, Section \ref{sec:prior}) is tailored for the air pollution applications, where the data $y|\bx$ exhibits positive skewness and heavy tail. To this end, our focus was to illustrate the importance of distribution calibration for the model's ability in uncertainty quantification, and have presented a practical approach that induces sufficient flexibility in capturing the non-Gaussian characteristics in the data, while maintaining analytical and computational tractability. However, it should also be recognized that the model space of \gls{HEX-GP} does not contain all the possible distribution patterns, e.g., negative skewness or extreme kurtosis. For applications where modeling such phenomenon is important, we note that it is straightforward to extend the recipe in (\ref{eq:hex_gp_skew}) to further complexity. For example, the negative skewness can be modeled by introducing double-exponential modification to the mean component $\mu(\bx)$ \citep{sager2019double}, and the extreme kurtosis can be modeled by introducing a inverse Gamma prior to $s(\bx)$. However, this added flexibility invariably introduces additional (nonparametric) parameters to the model, incurring a trade-off between the model's flexibility and its statistical / computational efficiency. To this end, we recommend practitioners to take a step-wise approach to model inference (similar to Section \ref{sec:inference}) to ensure the added complexity brings meaningful improvement to model performance.

Overall, we anticipate 
the benefits of adaptive weights that maximize predictive accuracy and, thus, minimize
exposure measurement error to be widely applicable in a variety of settings in which space-time prediction is of interest.   The approach yields rigorous uncertainty quantification that can be used to answer multiple relevant scientific questions, likely in multiple fields. We have found that this approach more effectively characterizes the sources of uncertainty associated with prediction of fine particulate matter concentrations in Eastern New England.

%% file: sections/table.tex
\begin{table}[ht]
    \caption{Summary of ensemble methods for simulation experiments in Section \ref{sec:exp}}
    \label{tab:exp_methods}
    
    \adjustbox{max width=\textwidth}{
    \centering
    \fbox{
    \begin{tabular}{|l|c|c|c|c|}
      Model Name & Label & Adaptive Ensemble & Bias Correction & Distribution Calibration \\
      \hline
      Bayesian Stacking & \textbf{Stacking} & & &
      \\
      Generalized Additive Ensemble & \textbf{GAM} &  & Yes &
      \\
      Adaptive Bayesian Model Averaging  & \textbf{BMA} & Yes &  &
      \\
      Bayesian Adaptive Ensemble & \textbf{BAE} & Yes & Yes &
      \\
      Bayesian Nonparametric Ensemble & \textbf{BNE} & Yes & Yes & Yes
    \end{tabular}
    }}
\end{table}

\begin{table}
\caption{Mean and Standard Deviation for RMSE and ECE in 1D time-series regression task.}
\label{tb:exp_1d_res}
\centering

\resizebox{0.95\columnwidth}{!}{%
\begin{tabular}{|c|c|c|c|c|c|c|c|c|c|}
\hline\hline
Metric & Model & $n=10$ & $n=25$ & $n=50$ & $n=100$ & $n=250$ & $n=500$ & $n=750$  & $n=1000$ \\
\hline
\multicolumn{10}{|c|}{Prediction Metrics}
\\
\hline
\multirow{5}{*}{MSE}
 &\textbf{Stacking} 
 & 0.5740 & 0.5703 & 0.4887 & 0.4742 & 0.4398 & 0.4262 & 0.4158 &
0.4314 \\
 &\textbf{GAM}
 & 0.4881 & 0.4344 & 0.4048 & 0.3233 & 0.2147 & 0.1939 & 0.1978 & 0.1733
\\
 &\textbf{BMA}
& 0.4195 & 0.4159 & 0.3894 & 0.3548 & 0.3399 & 0.3246 & 0.3085 & 0.3109
\\
&\textbf{BAE}
& 0.2562 & \textbf{0.2354} & 0.1917 & 0.1464 & 0.1086 & 0.0757 & 0.0657 & 0.0561
\\
&\textbf{BNE}
&\textbf{ 0.2492} & 0.2439 & \textbf{0.1816} & \textbf{0.1370} & \textbf{0.1029} & \textbf{0.0713} & \textbf{0.0606} & \textbf{0.0507}
\\ 
\hline
\multicolumn{10}{|c|}{Calibration Metrics}
\\
\hline
\multirow{5}{*}{NLL}
 &\textbf{Stacking} 
 & 1.1658 & 1.1509  & 1.1281 & 1.0855 & 1.0434 & 1.0307 & 1.0437 & 1.0103 
 \\
 &\textbf{GAM}
 & 1.5454 & 1.6761  & 1.7617 & 1.6693 & 1.5360 & 1.6702 & 1.7415 & 1.6142 
 \\
 &\textbf{BMA}
& 0.8706 & 0.8508 & 0.8474 & 0.8048 & 0.8031 & 0.7761 & 0.7751 & 0.7814 
\\
 &\textbf{BAE}
& 0.7489 & 0.7251 & 0.7030 & 0.6557 & 0.6603 & 0.6335 & 0.6417 & 0.6410 
\\
 &\textbf{BNE}
& \textbf{0.5545} & \textbf{0.5103} & \textbf{0.4597} & \textbf{0.3853} & \textbf{0.2765} & \textbf{0.2386} & \textbf{0.2256} & \textbf{0.2238} \\
\hline
\multicolumn{10}{|c|}{Coverage Probability Metrics}
\\
\hline
\multirow{5}{*}{ECE} 
& \textbf{Stacking} 
 & 0.0067 &  0.0063 & 0.0062 & 0.0063 & 0.0062 & 0.0060 & 0.0061 & 0.0064
\\
 &\textbf{GAM}
 & 0.0052 & 0.0048 & 0.0052 & 0.0049 & 0.0049 & 0.0042 & 0.0043 & 0.0045
\\
 &\textbf{BMA}
& 0.0064 & 0.0072 & 0.0070 & 0.0067 & 0.0075 & 0.0075 & 0.0075 & 0.0075
\\
 &\textbf{BAE}
& 0.0040 & 0.0046 & 0.0045 & 0.0042 & 0.0049 & 0.0051 & 0.0055 & 0.0052
\\
 &\textbf{BNE}
& \textbf{0.0023} & \textbf{0.0023} & \textbf{0.0020} & \textbf{0.0017} & \textbf{0.0013} & \textbf{0.0014} & \textbf{0.0015} & \textbf{0.0014}
\\ 
\hline
\end{tabular}
}
\end{table}

\newpage

\begin{table}
\caption{Mean and Standard Deviation for RMSE and ECE in 2D spatial regression task.}
\label{tb:exp_2d_res}
\centering

\resizebox{0.95\columnwidth}{!}{%
\begin{tabular}{|c|c|c|c|c|c|c|c|c|}
\hline\hline
Metric & Model & $n=25$ & $n=50$ & $n=100$ & $n=250$ & $n=500$ & $n=750$  & $n=1000$ \\
\hline
\multicolumn{9}{|c|}{Prediction Metrics} \\
\hline
\multirow{5}{*}{MSE}
 &\textbf{Stacking} 
& 0.7744  & 0.7357 & 0.7129 & 0.6947 & 0.6837 & 0.6769 & 0.6743 
\\
 &\textbf{GAM}
 & 1.8326 & 0.5341  & 0.2567 & \textbf{0.0915} & \textbf{0.0563} & \textbf{0.0496} & \textbf{0.0483}
\\
 &\textbf{BMA}
 & 0.2087 & 0.2008 & 0.1887 & 0.1725 & 0.1607 & 0.1561 & 0.1569
\\
 &\textbf{BAE}
 & \textbf{0.1558} & \textbf{0.1480} & \textbf{0.1379} & {0.1250} & {0.1164} & {0.1137} & {0.1136}
\\
 &\textbf{BNE}
 & 0.1566 & 0.1498 & 0.1397 & 0.1266 & 0.1189 & 0.1173 & 0.1169
\\ 
\hline
\multicolumn{9}{|c|}{Calibration Metrics} \\
\hline
\multirow{5}{*}{NLL}
 &\textbf{Stacking} 
 & -1.7356  & -1.7605 & -1.7815 & -1.7994 & -1.8106 & -1.8151 & -1.8188
\\
 &\textbf{GAM}
 & -1.0028 & -1.1653  & -1.2733 & -1.5408 & -1.7205 & -1.7676 & -1.7639
\\
 &\textbf{BMA}
 & -2.1288 & -2.1632 & -2.1988 & -2.2587 & -2.2914 & -2.3066 & -2.2892
\\
 &\textbf{BAE}
 & -2.4281 & -2.4555 & -2.4836 & -2.5292 & -2.5582 & -2.5720 & -2.5694
\\
 &\textbf{BNE}
 & \textbf{-2.5229} & \textbf{-2.5466} & \textbf{-2.5772} & \textbf{-2.6235} & \textbf{-2.6643} & \textbf{-2.6799} & \textbf{-2.6841}
\\ 
\hline
\multicolumn{9}{|c|}{Coverage Probability Metrics} \\
\hline
\multirow{5}{*}{ECE} 
& \textbf{Stacking} 
 & 0.0084 & 0.0089 & 0.0086 & 0.0084 & 0.0083 & 0.0085 &  0.0084 
\\
 &\textbf{GAM}
 & 0.0172 & 0.0179 & 0.0171 & 0.0183 & 0.0191 & 0.0185 & 0.0185
\\
 &\textbf{BMA}
 & 0.0132 & 0.0131 & 0.0125 & 0.0121 & 0.0115 & 0.0114 & 0.0117
\\
 &\textbf{BAE}
 & 0.0056 & 0.0054 & 0.0051 & 0.0046 & 0.0043 & 0.0041 & 0.0042
\\
 &\textbf{BNE}
 & \textbf{0.0044} & \textbf{0.0042} & \textbf{0.0039} & \textbf{0.0036} & \textbf{0.0034} & \textbf{0.0033} & \textbf{0.0035}
\\ 
\hline
\end{tabular}
}
\end{table}

\newpage


\begin{table}[ht]
\caption{Mean and Standard Deviation for spatial 6-fold cross validation \glsfirst{RMSE}, \glsfirst{NLL} and coverage probability of $95\%$ \glsfirst{CI}
for 2011 PM$_{2.5}$ predictions in Eastern New England, USA.}
\label{tb:app_comp}
\centering
\begin{adjustbox}{max width=\linewidth}
\begin{tabular}{|c|c|c|c|c|c|}
\hline\hline
\multirow{3}{*}{Model/Metric} & \multirow{3}{*}{\makecell{\textbf{RMSE}(mean)}} & \multirow{3}{*}{\textbf{RMSE}(median)}  & \multirow{3}{*}{\textbf{NLL}(mean)}  & \multirow{3}{*}{\textbf{NLL}(median)} & 
\makecell{$95\%$ CI \\ Coverage \\ Probability}
 \\  \hline

\textbf{Stacking} & $0.862 \pm 0.292$ & $0.785 \pm 0.292$ & $0.941 \pm 0.088$ & $0.914 \pm 0.088$ & $0.9412$ \\ 
\hline 
\textbf{GAM} & $1.096 \pm 0.161$ & $1.116 \pm 0.161$ & $1.208 \pm  0.409$ & $1.008 \pm 0.409$ & $0.9216$ \\
\hline 
\textbf{BMA} & $0.888 \pm 0.423$ & $0.842 \pm 0.423$ & $1.081 \pm  0.345$ & $0.930 \pm 0.345$ & $0.9412$ \\
\hline 
\makecell{\textbf{BAE}  \\ \scriptsize{(Prior Mean=Zero)}} & $ 0.872 \pm 0.397$ & $0.767 \pm 0.397$ & $1.051 \pm  0.223$ & $0.905 \pm 0.223$ & $0.9412$ \\
\hline 
\makecell{\textbf{BAE} \\ \scriptsize{(Prior Mean=Stacking)}} & $0.841 \pm 0.348$ & $0.719 \pm 0.348$  & $0.967 \pm 0.137$ & $0.897 \pm 0.137$ & $0.9412$ \\ 
\hline 
\makecell{\textbf{BNE} \\ \scriptsize{(Prior Mean=Zero)}} & $0.872 \pm 0.397$ & $0.775 \pm 0.397$  & $1.053 \pm  0.212$ & $0.928 \pm  0.212$ & $0.9412$\\ 
\hline  
\makecell{\textbf{BNE} \\ \scriptsize{(Prior Mean=Stacking)}} & $0.840 \pm 0.349$ & $0.724 \pm 0.349$  & $0.962 \pm  0.127$ & $0.892 \pm  0.127$ & $0.9608$\\
\hline 

\end{tabular}
\end{adjustbox}
\end{table}

%% file: sections/figure.tex
\clearpage

\begin{figure}[ht]
    \centering
    \includegraphics[
    width=0.55\linewidth, height=0.4\linewidth]{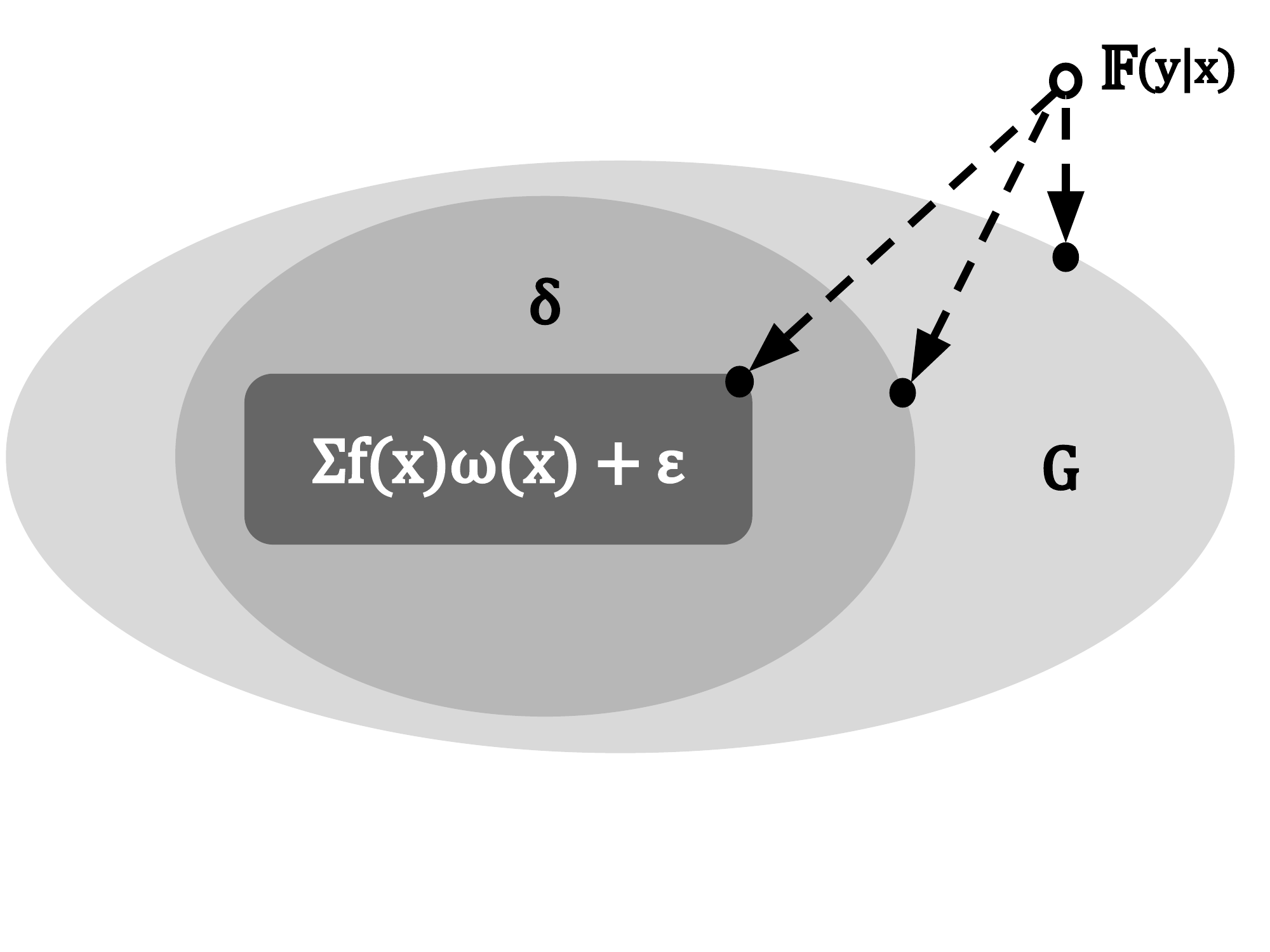}
    \vspace{-3em}
    \caption[caption]{Illustration of the BNE's model space.
    \textbf{Dark Gray}: $\Msc_{\omega}$; \textbf{Medium Gray}: $\Msc_{\omega, \delta}$; \textbf{Light Gray}: $\Msc_{\omega, \delta, G}$;
    \textbf{Empty Dot}: Empirical Distribution $\Fbb(y|\bx)$; 
    \textbf{Black Dots}: Projection of $\Fbb(y|\bx)$ to Model Spaces.}
    \label{fig:model_space}
\end{figure}

\begin{figure}
\centering
\includegraphics[trim=80 20 80 20,clip,width=\linewidth]{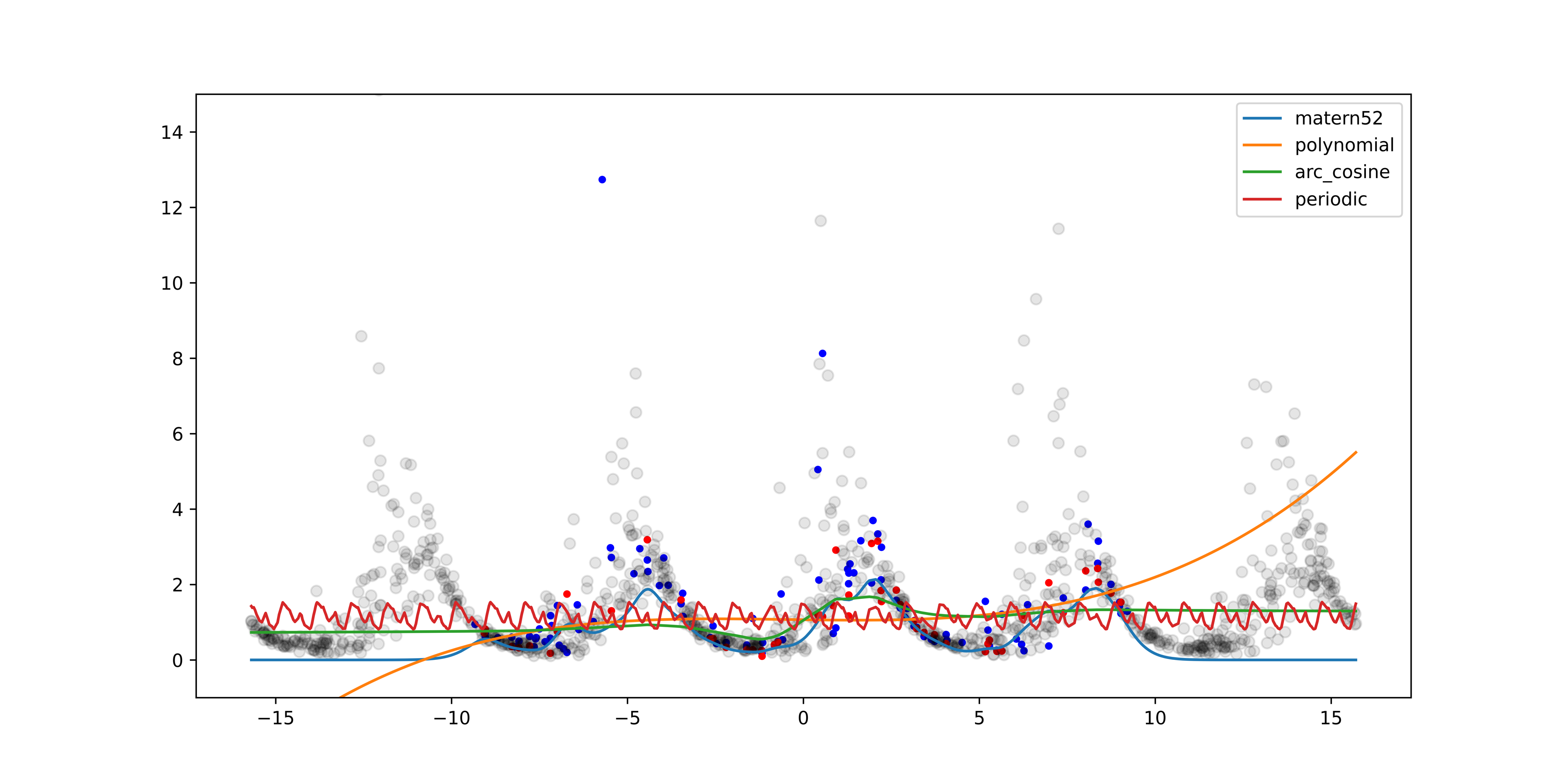}
  \caption{Samples from the data-generating distribution and the deterministic predictions from base models in the time-series experiment. }
	\label{fig:base_exp_1d}  
\end{figure}


\begin{figure}[!htb]
\centering
\resizebox{1.\columnwidth}{!}{%
\begin{tabular}[t]{@{\hspace{-5pt}}c@{\hspace{-5pt}}c@{\hspace{-5pt}}}
    \begin{tabular}{c}
    \subfloat{\includegraphics[trim=70 10 70 10,clip,width=\linewidth]{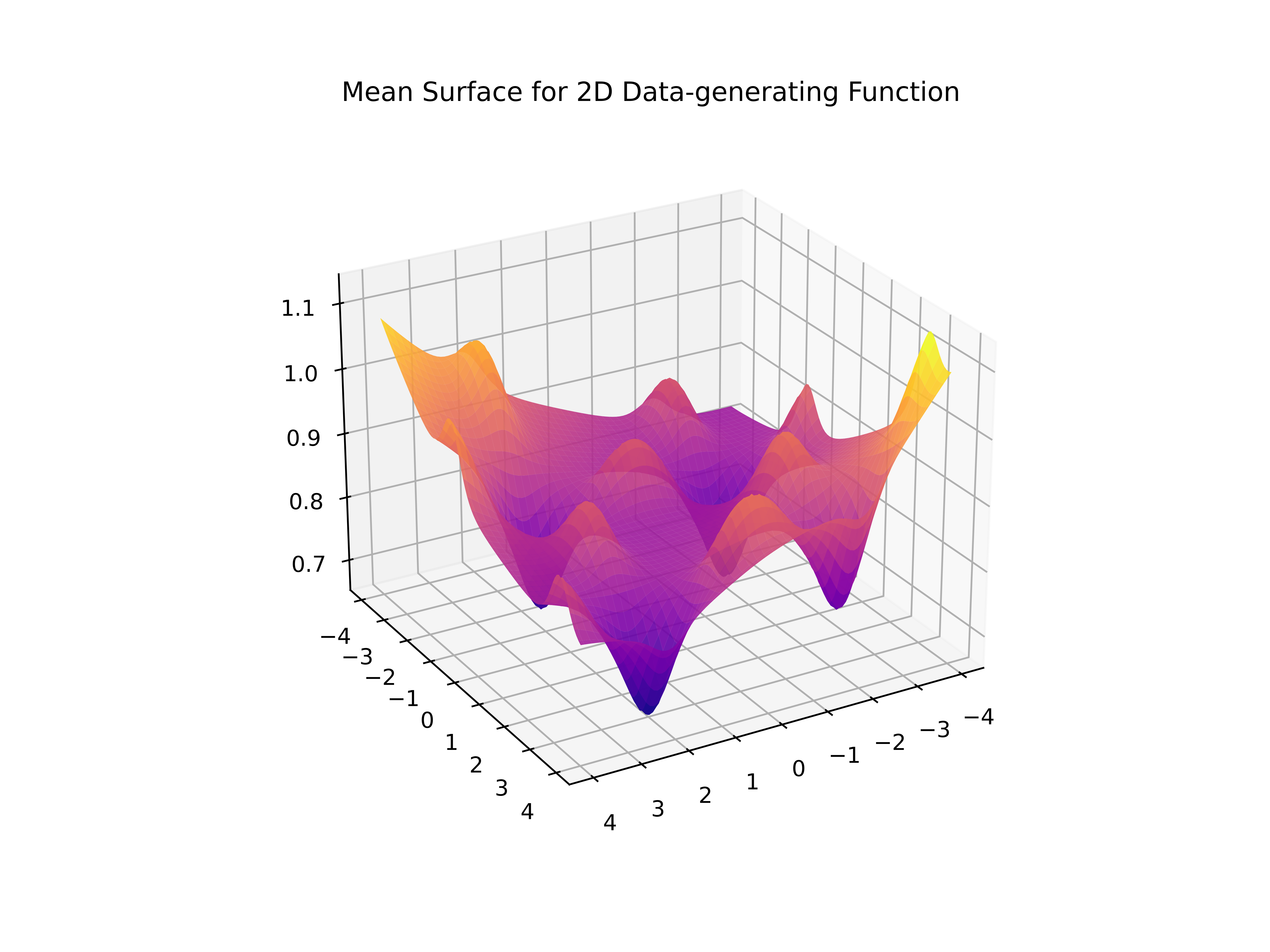}}
    \end{tabular}        
&
    \begin{tabular}{c}
    \subfloat{\includegraphics[trim=70 10 70 10,clip,width=\linewidth]{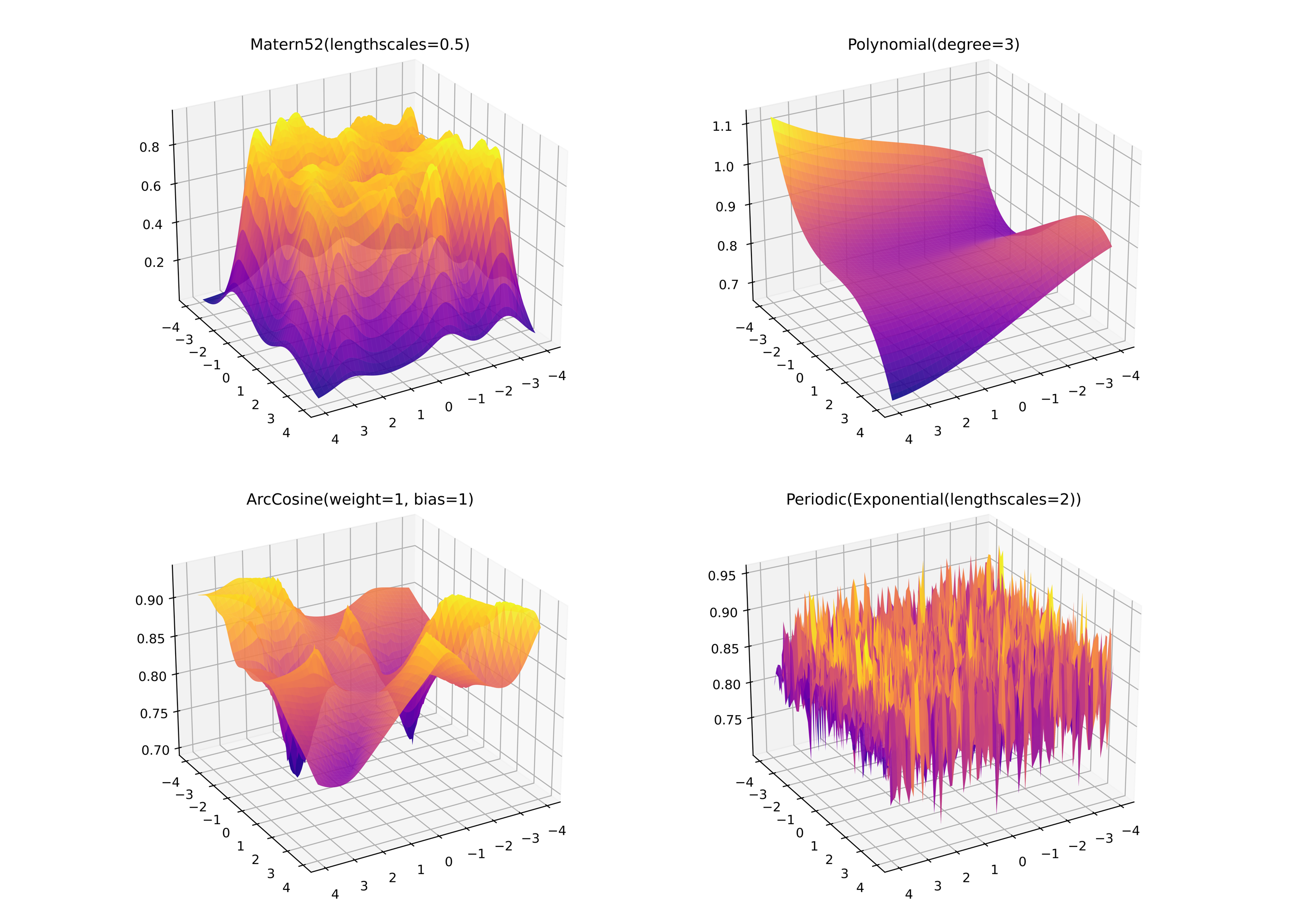}}
    \end{tabular}
\end{tabular}  
}
  \caption{Samples from the data-generating distribution (left) and the deterministic predictions (right) from base models in the spatial experiment. }
	\label{fig:base_exp_2d}  
\end{figure}

\newpage

\begin{figure}[ht]
    \centering
    
    \includegraphics[width=.32\linewidth]{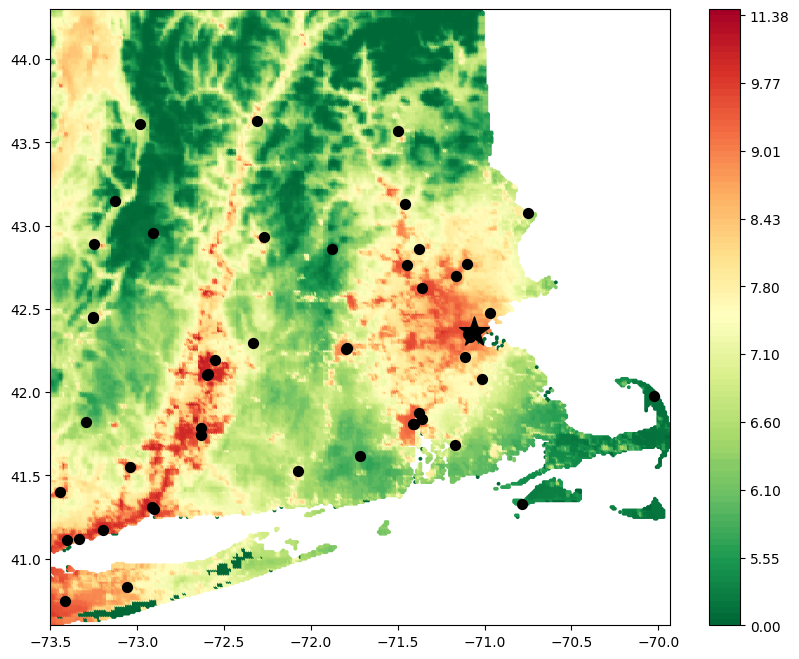}
    \includegraphics[width=.32\linewidth]{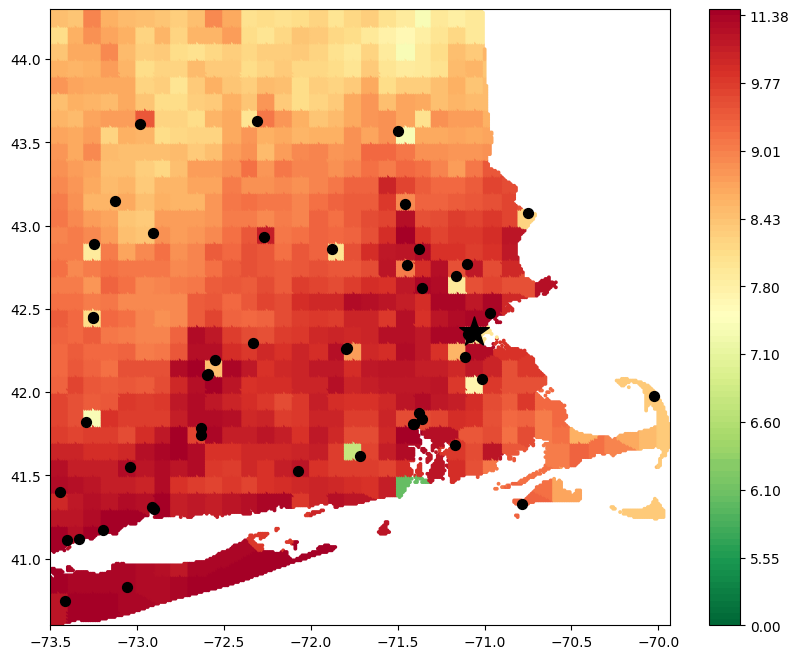}
    \includegraphics[width=.32\linewidth]{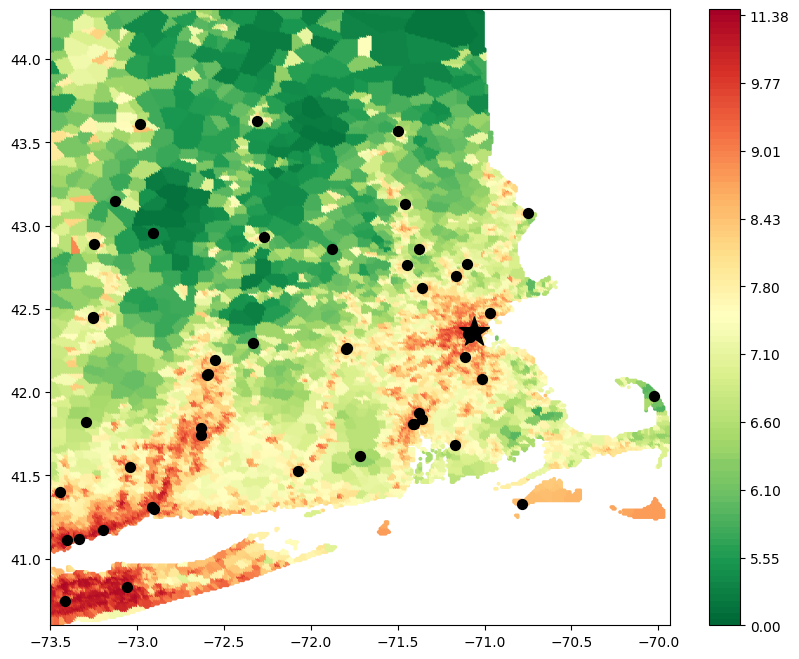}
    
    \includegraphics[width=.32\linewidth]{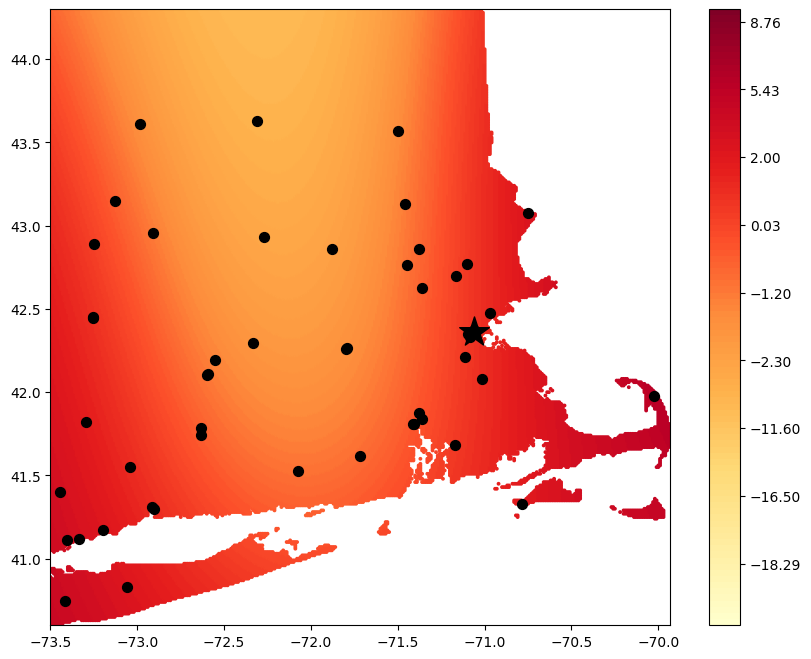}
    \includegraphics[width=.32\linewidth]{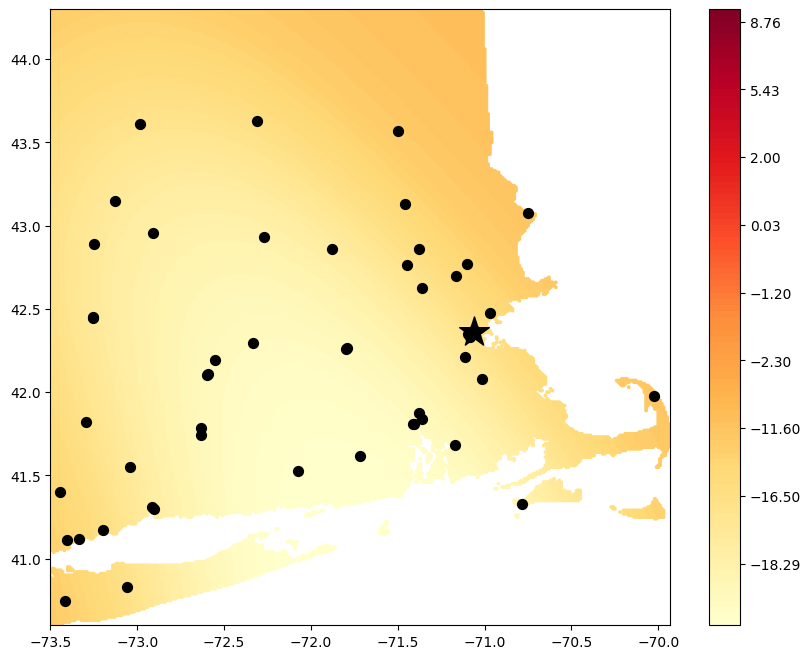}
    \includegraphics[width=.32\linewidth]{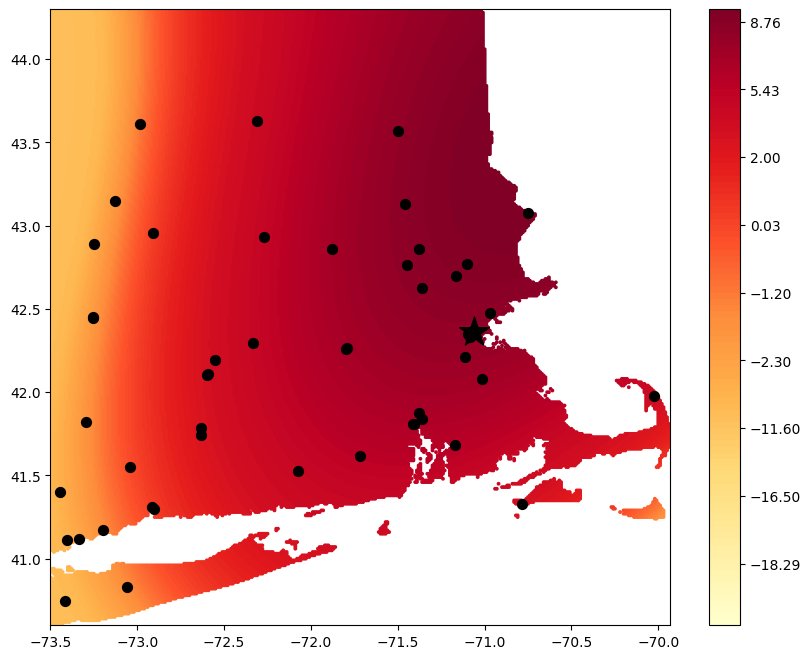}
    \caption{
    Annual average PM$_{2.5}$ predictions (Top) and posterior mean of Gaussian Processes (Bottom) assigned to each air pollution models in Eastern Massachusetts in 2011. 
    \textbf{Left}:MH2021 \citep{kloog_new_2014}; \textbf{Middle}:GS2018; 
    \textbf{Right}:SK2020; 
    \textbf{Black Dots}: Monitor Locations; \textbf{Black Star}: Boston, MA.}
    \label{fig:pm25_base}
\end{figure}

\newpage



\begin{figure}[ht]
    \centering
    \subfloat[Posterior Mean]{
    \includegraphics[width=0.4\textwidth]{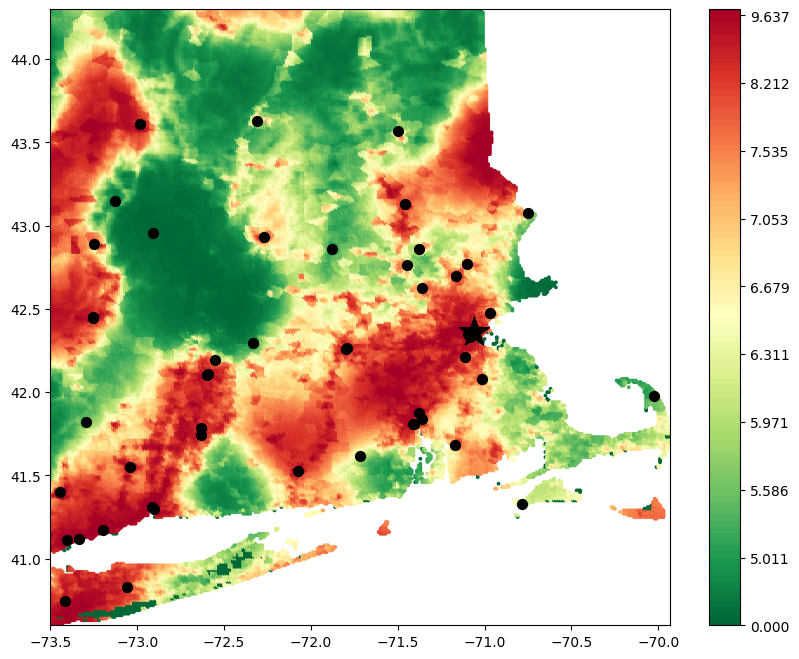}
    \label{fig:app_post_mean}
    }
    \subfloat[Overall Uncertainty]{
    \includegraphics[width=0.4\textwidth]{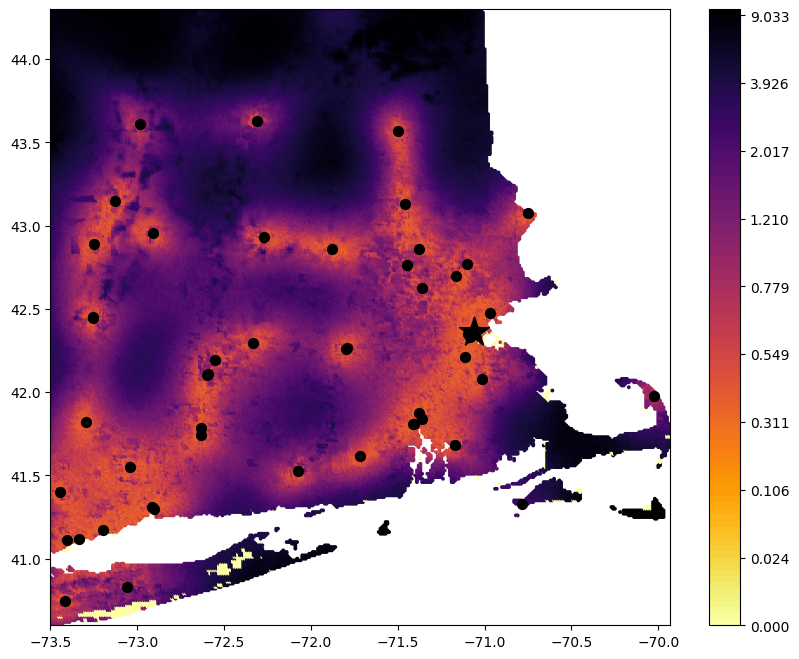}
    \label{fig:app_post_unc}
    }

    \subfloat[Uncertainty, Model Selection]{
    \includegraphics[width=0.4\textwidth]{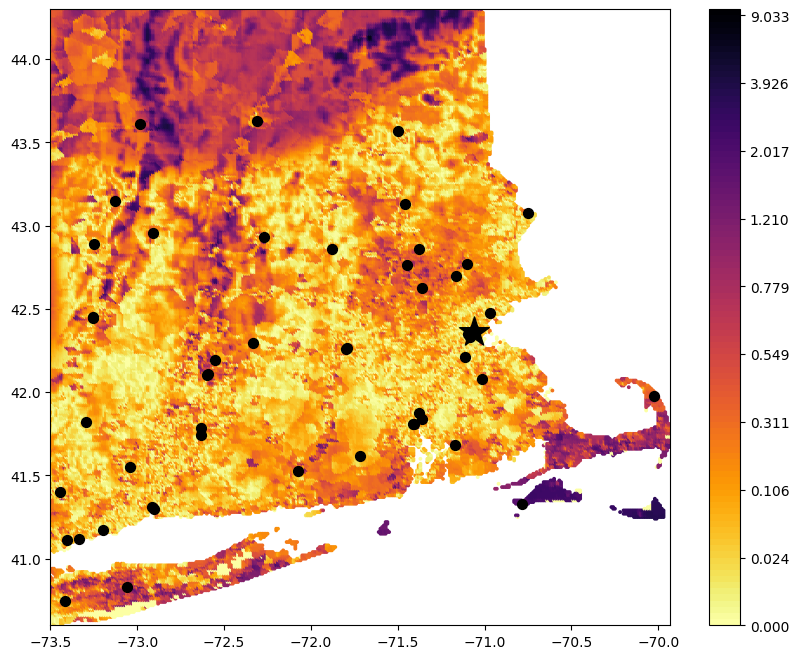}
    \label{fig:app_post_uncn_ens}
    }
    \subfloat[Uncertainty, Prediction]{
    \includegraphics[width=0.4\textwidth]{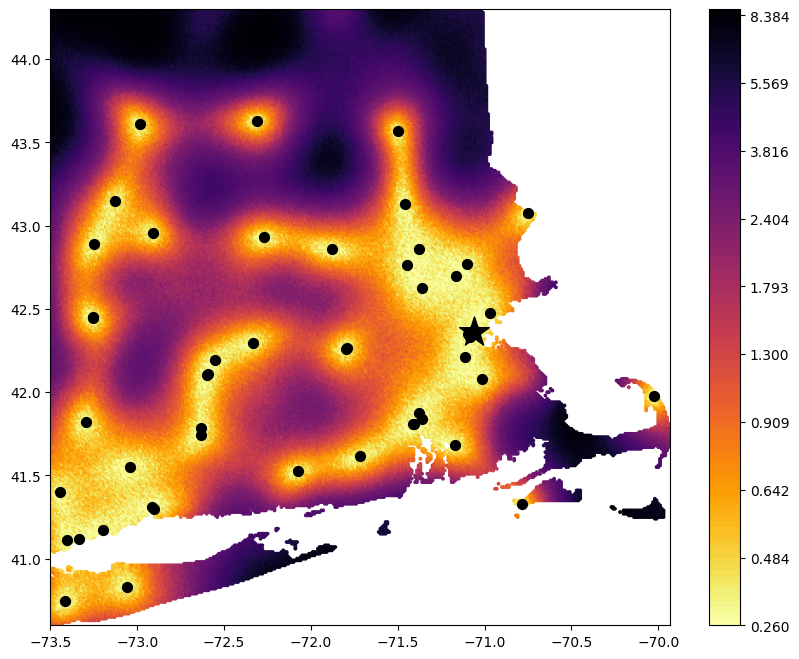}
    \label{fig:app_post_uncn_resid}
    }
    \caption{
    Posterior predictive mean, predictive uncertainty (i.e. variance) and its decomposition in the BNE model. \textbf{Black Dots}: Location of air pollution monitors; \textbf{Black Star}: Boston, MA.}
    \label{fig:app_res}
\end{figure}

%% file: sections/appendix_v1.tex


\input{"./sections/exp_supp"}

\section{Scalable Gaussian Process Computation via Random-feature Parameterization}
\label{sec:rff}

\input{"./sections/rff"}


\input{"./sections/property_v1"}


\subsection{Proof for Theorem \ref{thm:accuracy_rmse}}
\label{sec:accuracy_proof}
\input{"./sections/proof/bne_accuracy"}

%% file: sections/exp_supp.tex
\section{Data Generation Mechanism}
\label{sec:exp_spp_data_gen}

To prepare the base models for the ensemble methods, we specify a library of base models $\Fsc=\{\hat{f}_k\}^K_{k=1}$, where each $\hat{f}_k$ is pre-trained on simulated data of sample size 50 (for time-series experiments) or 250 (for spatial experiments). In this work, we choose each $f_k$ to be kernel ridge regression models with the following kernel functions: (1) a polynomial kernel with degree 3 $k(x_1, x_2)=(1+x_1^\top x_2)^3$, (2) an arc-cosine kernel $k(x_1, x_2) = 1 - \frac{1}{\pi} arccos(\frac{x_1^\top x_2}{|x_1||x_2|})$ whose model space is equivalent to that of a 1-layer neural network with infinite-number of neurons \citep{cho2009kernel},  (3) an  exponential kernel $k(x_1, x_2) = exp(-|x_1 - x_2|/l)$ whose model space represents the space of continuous (and possibly non-differentiable) functions,  and (4) Mat\'{e}rn $\frac{5}{2}$ kernel whose model space represents the space of continuous functions that are at least twice differentiable \citep{williams2006gaussian}. 
As a result, the library $\Fsc=\{f_k\}_{k=1}^4$ as a whole encodes a diverse set of assumptions about the data generating function $f(\bx)$, but none of the individual models $f_k$ can predict $f(\bx)$ universally well across the input space $\Xsc$, especially in locations without training data (see Figure 3 in the main text). ~
Then, using $\Fsc$, we fit each ensemble method in Table 1 in the main text 
on simulated data of sample size $n \in (50, 1000)$, and we compute each model's operating characteristics with respect to prediction (\gls{RMSE}) and  uncertainty quantification (\gls{NLL} and \gls{ECE}) on a separate validation dataset with sample size $1000$.
We repeat the simulation for each model-sample size combination 100 times. 

%% file: sections/rff.tex
As discussed in Section 4 in the main text
, the computation of the Gaussian process posterior often involves inverting a $n \times n$ kernel matrix over the entire dataset, leading to a $O(n^3)$ computational complexity and making it infeasible for large-scale spatio-temporal problems. To this end, \citep{rahimi_random_2008} has proposed a random-feature based parameterization of the Gaussian process prior that effectively reduced \gls{GP} to a Bayesian linear model, hence allowing for scalable $O(n)$ computation when combined with a gradient-based MCMC procedure such as \gls{NUTS} \citep{hoffman_no-u-turn_2011} or the \gls{SGLD} \citep{welling2011bayesian}.

Specifically, consider a Gaussian process prior over data $\{\bx_i\}_{i=1}^n$: $\textbf{f} \sim MVN(0, \bK_{n \times n})$, where $\bK_{i,j} = k(\bx_i - \bx_j)$ is a $n \times n$ matrix for a shift invariant (e.g., the RBF kernel $k_{RBF}(\bx_i - \bx_j)=exp(-||\bx_i - \bx_j||^2_2/l)$). The random-feature method shows that a \gls{GP} prior can be equivalently written in a linear model form:
\begin{align}
    \textbf{f} = \bH_{n \times D}\bbeta_{D \times 1}, \quad \bbeta_{D \times 1} \sim MVN(\bzero, \bI_{D \times D}),
    \label{eq:rfgp}
\end{align}
where $\bH_{n \times D} = \frac{1}{\sqrt{D}}[h(\bx_1)_{D \times 1}, \dots, h(\bx_n)_{D \times 1}]^\top$ is the \textit{random-feature} transformation of the data $\textbf{\bx}$.  Note that (\ref{eq:rfgp}) can be equivalent written as:
$$f(\bx) \sim MVN(\bzero, \hat{\bK} = \bH\bH^\top),$$ 
where 
\begin{align}
    \hat{\bK}_{i, j} = \hat{k}(\bx_i - \bx_j) = \frac{1}{D} h(\bx_j)^\top h(\bx_j) = \frac{1}{D} \sum_{p=1}^D h_p(\bx_i)h_p(\bx_j) = \hat{E}_p(h_p(\bx_i)  h_p(\bx_j)).
    \label{eq:rf_approx}
\end{align}
is the approximated kernel function.

The transformation function $h: \real^d \rightarrow \real^D$ takes the below form:
\begin{align}
h(\bx) = \sqrt{2} * cos(\bW_{D \times d}\bx_{d \times 1} + \bb_{D \times 1})
\label{eq:rff}
\end{align}
where elements of $\bb$ are drawn i.i.d. from $Unif(0, 2\pi)$. The elements of $\bW_{D \times d}$ are drawn from $f(w)$, which is the Fourier transform of the shift-invariant kernel $k(\delta)$ for $\delta=\bx - \by$. As a result, the random-feature formulation of (\ref{eq:rff}) leads to a strong approximation guarantee of the \gls{GP} kernel $k(\bx_i - \bx_j)$. That is, by the classic Bohner theorem, any shift-invariant kernel $k$ can be expressed as the Fourier transformation \citep{williams2006gaussian}:
\begin{align*}
    k(\bx_i - \bx_j) = Re\Big( \int_{w \in \real^d} exp(-z w^\top(\bx_i - \bx_j))  f(w) dw \Big) = E_w(h_w(\bx_i)^\top h_w(\bx_j))
\end{align*}
where $z$ is the imaginary unit and $h_w(\bx)=cos(w^\top\bx+b)$ for $w \sim f(w)$ and $b \sim Unif(0, 2\pi)$. This shows that (\ref{eq:rf_approx}) is indeed a valid, Monte-carlo-based approximation of the true kernel function $k$ with a fast exponential rate of convergence in terms of the number of random features $D$ \citep{rahimi_random_2008, liu2020random}.

Consequently, to approximate a Gaussian process model with shift-invariant kernel $k(\delta)$, one only need to derive its Fourier density $f(w)$ and construct a $D$-dimensional linear model following (\ref{eq:rfgp}) and (\ref{eq:rff}). Many classic kernels $k$ correspond to well-known distributions $f$. For example, the \gls{RBF} kernel corresponds to Gaussian distribution, while the Mat\'{e}rn kernel corresponds to Student's t-distribution \citep{choromanski2018geometry}. Due to its simplicity and strong empirical performance, the random feature method has seen widespread adoption in both the applied and theoretical machine learning literature for building large-scale, high-dimensional probabilistic models
\citep{liu2020random, cutajar2017random, liu2020simple, ghorbani2021linearized, mei2019generalization, mei2021learning}.

Coming back to the \gls{BNE} model, recall the full BNE model likelihood presented in Section 4 of the main text:
\begin{alignat*}{2}
y|\bx &\sim N(\sum_k \bomega_k(\bx)f(\bx) + &&\delta(\bx) + m(\bx), \sigma^2 * s^2(\bx)) \nonumber \\
\bomega &= \frac{exp(w_k)}{\sum_k exp(w_k)}, \quad w_k &&\sim GP(0, k_w ) \\
\delta &\sim GP(0, k_\delta), \quad log \, s && \sim GP(0, \sigma_s^2 * k_s ) \\
m &\sim Exp(\lambda), \quad log \, \lambda && \sim GP(\mu_\lambda, \sigma^2_\lambda * k_\lambda).
\end{alignat*}
Using the random-feature parametrization, the GP priors in the original BNE model are reduced to:
\begin{alignat*}{2}
w_k(\bx) &= h(\bx)^\top\bbeta_k, 
\quad &&\bbeta_{k, D \times 1} \sim MVN(0, \bI)\\
\delta(\bx) &= h(\bx)^\top\bbeta_\delta, 
\quad &&\bbeta_{\delta, D \times 1} \sim MVN(0, \bI)\\
log \, s(\bx) &= h(\bx)^\top\bbeta_s,
\quad &&\bbeta_{s, D \times 1} \sim MVN(0, \sigma_s^2\bI)\\
log \, \lambda(\bx) &= h(\bx)^\top\bbeta_\lambda + \mu_\lambda,
\quad &&\bbeta_{\lambda, D \times 1} \sim MVN(0, \sigma_\lambda^2\bI),
\end{alignat*}
As a result, we only need to perform posterior inference with respect to K+3 Bayesian linear models $(\{\beta_k\}_{k=1}^K, \beta_\delta, \bbeta_s, \bbeta_\lambda)$ with fixed dimensions $D$, reducing the computation complexity from $O(n^3)$ to $O(n)$. 
Furthermore, due to the low-dimensionality of spatio-temporal modeling (i.e., d=3), a small amount of random features (e.g., $D=128$) is already sufficient of obtaining high-quality approximation to the GP posterior.

%% file: sections/property_v1.tex
\section{Theoretical Properties}
\label{sec:theory}

\subsection{Prediction}
\label{sec:bne_prediction}

Recall that the goal of a \gls{BNE} is to produce both calibrated predictive uncertainty and accurate prediction.  
Here, we show in Theorem \ref{thm:accuracy_rmse} that \textit{calibration preserves accuracy}. That is, when compared to an uncalibrated model (i.e. the model estimated only using Step 1 and 2), the calibrated model's excess predictive error in terms of the \gls{MSE} is bounded and asymptotically approaches zero. Intuitively, this result implies that for reasonably large sample sizes, the calibrated model cannot have worse predictive accuracy than the uncalibrated model.

\begin{theorem}[Calibration Preserves Accuracy, \gls{MSE} Loss]
Denote $F_\mu \in \Fsc$ the uncalibrated predictive CDF, and $F=G \big[ F_\mu \big] \in \Fsc$ the calibrated predictive CDF. 
Given observations $Y=\{y_i\}_{i=1}^N$ generated from  $F^*(y|\bx_i)'s$, we denote $\bbL(Y,F) = \frac{1}{N} \sum_{i=1}^n (y_i - E(y|\bx_i))^2$ the \gls{MSE} loss, and denote $\bbC(Y, F)$ the calibration loss:
\begin{align*}
\bbC(Y, \hat{F}) &= \int_{y \in \Ysc}
\frac{1}{N}\sum_{i=1}^n 
\Big| \bbI(y_i < y|\bx_i) - F(y|\bx_i) \Big| dy.
\end{align*}
i.e. the empirical $L_1$ loss between model and empirical \gls{CDF}s.

Assume that $F$ is a consistent estimator for $F^*$, i.e. $\underset{n \rightarrow \infty}{limsup}  \;\bbC(Y, \hat{F}) \leq \epsilon_n$ for a sequence  $\epsilon_n \rightarrow 0$. The excess error of the calibrated CDF $F$ with respect to the uncalibrated CDF $F_\mu$ in terms of the \gls{MSE} loss $\bbL$ is bounded by the asymptotically vanishing term $\epsilon_n$ up to a constant $C$:
\begin{align}
\underset{n \rightarrow \infty}{limsup} \Big[
\bbL(Y, \hat{F}) - \bbL(Y, \hat{F}_0) 
\Big] \leq C \epsilon_n
\label{eq:excess_risk_rmse}
\end{align}
\label{thm:accuracy_rmse}
\end{theorem}

\begin{proof}
See Section \ref{sec:accuracy_proof}
\end{proof}
Specifically, the expression in (\ref{eq:excess_risk_rmse}) suggests that a calibration model with faster convergence rate (i.e. smaller $\epsilon_n$) leads to greater improvement in the prediction performance (in the sense that the excess error is small). To this end, 
we notice that by using an Mat\'{e}rn $\frac{3}{2}$ kernel with the length-scale parameter selected adaptively from data, $\epsilon_n$ may reach the minimax-optimal rate up to a logarithmic factor \citep{van_der_vaart_adaptive_2009}, leading to improved prediction performance in finite samples.

%% file: sections/proof/bne_accuracy.tex
Theorem \ref{thm:accuracy_rmse} is a special case of Theorem \ref{thm:accuracy} when the accuracy measure is \gls{MSE}. We present Theorem \ref{thm:accuracy} below. 

\begin{theorem}[Calibration Preserves Accuracy, General Loss]
Denote $F_0 \in \Fsc$ the uncalibrated predictive CDF, and $F=G \circ F_0 \in \Fsc$ the calibrated predictive CDF. 
Given observations $\{y_i, \bx_i\}_{i=1}^N$, we denote:
\begin{enumerate}
\item $\bbL(Y,F) = \frac{1}{N} \sum_{i=1}^N L(y_i, F(Y|\bX=\bx_i))$ with respect to a accuracy measure function $L: \Ysc \times \Fsc \rightarrow \real$ that is bounded and proper, i.e.:
\begin{enumerate}
\item \textbf{Bounded}: There exists positive constant $B<\infty$ such that $L(y, F) < B \quad \forall y \in \Ysc, F \in \Fsc$. Furthermore, for $F, F' \in \Fsc$ such that $\int_{\Ysc} \Big|F(t) - F'(t)\Big| dt < \epsilon$, we have: 
$$\int_{\Ysc} \Big|L(y, F') - L(y, F)\Big| dy < B\epsilon.$$
\item \textbf{Proper}: $\bbL(Y, F)$ is minimized by the true CDF $\Fbb$, i.e. $\bbL(Y, \Fbb) \leq \bbL(Y, F) \quad \forall F \in \Fsc$.
\end{enumerate}
\item $\bbC(Y, \hat{F})$ the empirical calibration loss defined as:
\begin{align*}
\bbC(Y, \hat{F}) &= \int_{\Ysc}
\frac{1}{N}\sum_{i=1}^n 
\Big| \bbI(y_i < t|\bx_i) - F(t|\bx_i) \Big| dt
\end{align*}
\end{enumerate}
Furthermore, we assume that $\hat{F}$ is $\epsilon_n$-calibrated \citep{abernethy2011blackwell}, i.e. $$\underset{n \rightarrow \infty}{limsup}  \;\bbC(Y, \hat{F}) \leq \epsilon_n$$ for some $\epsilon_n > 0$. 

Then the excess error of the calibrated CDF $\hat{F}$ with respect to the uncalibrated CDF $\hat{F}_0$ in terms of $\bbL$ is bounded by $\bbC$, i.e.
\begin{align*}
\bbL(Y, \hat{F}) - \bbL(Y, \hat{F}_0) \leq 2B * \bbC(Y, \hat{F}) + (B + 1)*\epsilon_n
\end{align*}
Morever, since $\hat{F}$ is $\epsilon_n$-calibrated:
\begin{align}
\underset{n \rightarrow \infty}{limsup} \Big[
\bbL(Y, \hat{F}) - \bbL(Y, \hat{F}_0) 
\Big] \leq (3B + 1)*\epsilon_n
\label{eq:excess_risk}
\end{align}
\label{thm:accuracy}
\end{theorem}

Our proof technique is an adaptation of the game-theoretic analysis of internal regret from the setting of online sequential classification \citep{cesa-bianchi_prediction_2006, kuleshov2017estimating}, and does not assume independence between the observations.

\begin{proof} 
For $Y, \bX \sim \Fbb(Y, \bX)$, denote $l(Y, F)(t)= L(Y, F(t|\bX))$ the accuracy measure that involves the observation $Y$ and the corresponding predictive CDF $F(Y|\bX)$ evaluated at $t \in \Ysc$. For an observed data pair $\{y_i, \bx_i \}$, denote the corresponding empirical version of $l$ as $l_i(Y, F)(t)= L(y_i, F(t|\bx_i))$. Also denote $F_i(t)=F(t|\bX=\bx_i)$.

We notice below facts:
\begin{enumerate}
\item $l(Y, F)(t)$ is bounded by $B$, therefore for any predictive CDFs $F, F' \in \Fsc$, we always have:
$$l(Y, F)(t) - l_i(Y, F')(t) \leq B$$
\item Since $l(Y, F)(t)$ is proper in the sense that this loss is minimized by the true model parameters, then for $Y \sim \Fbb$:
$$\Fbb(t) \in
 \underset{F \in \Fsc}{argmin} \; 
E_{Y \sim \Fbb} \Big( l(Y, F)(t) \Big)
$$

\item $\bbI \Big( \hat{F}_i(t) = p_i \Big) = 
\bbI \Big( \hat{F}_i(t) = p_i, y_i < t \Big) + 
\bbI \Big( \hat{F}_i(t) = p_i, y_i > t \Big)$. \\
Furthermore,
\begin{align*}
\bbI \Big( \hat{F}_i(t) = p_i, y_i < t \Big) &=
\bbI \Big( \hat{F}_i(t) = p_i \Big) * 
\Big( \bbI \big( y_i < t | \hat{F}_i(t) = p_i \big) -  p_i \Big) + 
\bbI \Big( \hat{F}_i(t) = p_i \Big) * p_i
\\
\bbI \Big( \hat{F}_i(t) = p_i, y_i > t \Big) &=
\bbI \Big( \hat{F}_i(t) = p_i \Big) * 
\Big( p_i - \bbI \big( y_i < t | \hat{F}_i(t) = p_i \big)\Big) + 
\bbI \Big( \hat{F}_i(t) = p_i \Big) * 
\Big( 1 - p_i \Big)
\end{align*}
which implies:
$$\bbI \Big( \hat{F}_i(t) = p_i \Big) 
\leq 
2 * \bbI \Big( \hat{F}_i(t) = p_i \Big) * 
\Big| \bbI \big( y_i < t | \hat{F}_i(t) = p_i \big) -  p \Big| + 
\bbI \Big( \hat{F}_i(t) = p_i \Big) * p_i(t)
$$
where $p_i(t) = p_i$ if $y_i < t$ and $p_i(t) = 1 - p_i$ otherwise.
\end{enumerate}
Combining above facts, we have:
\begin{align}
& \bbL(Y, \hat{F})(t) - \bbL(Y, \hat{F}_0)(t) \nonumber\\
&=
\frac{1}{N} \sum_{i=1}^N
\Big( l_i(Y, \hat{F})(t) - l_i(Y, \hat{F}_0)(t) \Big) * \bbI(\hat{F}_i(t)=p_i) 
\nonumber \\
&\leq 2 \frac{1}{N} \sum_{i=1}^N 
\underline{\Big( l_i(Y, \hat{F})(t) - l_i(Y, \hat{F}_0)(t) \Big)} * 
\Big| \bbI \big( y_i < t | \hat{F}_{i}(t) = p_i \big) -  p_i \Big|  * 
\bbI(\hat{F}_i(t)=p_i)  + 
\nonumber \\
& \qquad \frac{1}{N}\sum_{i=1}^N  
\Big( l_i(Y, \hat{F})(t) - l_i(Y, \hat{F}_0)(t) \Big) * 
p_i(t) * 
\bbI(\hat{F}_i(t)=p_i) 
\nonumber \\
&\leq 2B * \bigg\{
\frac{1}{N} \sum_{i=1}^N 
\Big| \bbI \big( y_i < t | \hat{F}_{i}(t) = p_i \big) -  p_i \Big| * 
\bbI(\hat{F}_i(t)=p_i) 
\bigg\}
\; + 
\nonumber \\
& \qquad \bigg\{ 
\frac{1}{N}\sum_{i=1}^N  
\Big( l_i(Y, \hat{F})(t) - l_i(Y, \hat{F}_0)(t) \Big) * 
p_i(t) * 
\bbI(\hat{F}_i(t)=p_i) \bigg\}
\label{eq:proof_part1}
\end{align}
Here the first inequality uses Fact 3. The second inequality uses  Fact 1 on the underlined component.

We now consider the two expressions in curly brackets in (\ref{eq:proof_part1}). Notice the first expression corresponds to an "unintegrated" version of calibration loss:
\begin{align}
\frac{1}{N} \sum_{i=1}^N 
\Big| \bbI \big( y_i < t | \hat{F}_{i}(t) = p_i \big) -  p_i \Big| * 
\bbI(\hat{F}_i(t)=p_i) 
& = 
 \frac{1}{N} \sum_{i=1}^N 
\Big| \bbI \big( y_i < t | \bx_i \big) -  \hat{F}_{i}(t) \Big| 
= \bbC(Y, \hat{F})(t),
\label{eq:exp_1}
\end{align}
and the second expression can be written as:
\begin{align}
& \quad \frac{1}{N}\sum_{i=1}^N  
\Big( l_i(Y, \hat{F})(t) - l_i(Y, \hat{F}_0)(t) \Big) * 
p_i(t) * \bbI(\hat{F}_i(t)=p_i) 
\nonumber\\
& =
\frac{1}{N} \left[\begin{aligned}
&\sum_{y_i < t}
\Big( l_i(Y, \hat{F})(t) - l_i(Y, \hat{F}_0)(t) \Big) * \hat{F}_i(t) \\
&{}+ \sum_{y_i \geq t}
\Big( l_i(Y, \hat{F})(t) - l_i(Y, \hat{F}_0)(t) \Big) * (1 - \hat{F}_i(t))
\end{aligned}\right]
\nonumber\\
& \leq
\frac{1}{N} \left[\begin{aligned}
&\sum_{y_i < t}
\Big( l_i(Y, \hat{F})(t) - l_i(Y, \hat{F}_0)(t) \Big) * \bbI\big( y_i < t | \bx_i \big) \\
&{}+ \sum_{y_i \geq t}
\Big( l_i(Y, \hat{F})(t) - l_i(Y, \hat{F}_0)(t) \Big) * \bbI\big( y_i > t | \bx_i \big)
\end{aligned}\right] + \epsilon_n
\nonumber\\
& = \mathbb{E}
\Big( l(Y, \hat{F})(t) - l(Y, \hat{F}_0)(t) \Big) + \epsilon_n 
\nonumber\\
& \leq 
\mathbb{E}
\Big( l(Y, \Fbb)(t) - l(Y, \hat{F}_0)(t) \Big) + (B + 1)\epsilon_n 
\nonumber\\
& \leq (B + 1)\epsilon_n
\label{eq:exp_2}
\end{align}
where the first inequality follows since $\hat{F}$ is $\epsilon_n$-calibrated, the second inequality follows since $l$ is a bounded loss, and the last inequality follows since $l$ is a proper loss.

Now replace the two expressions within brackets in (\ref{eq:proof_part1}) with (\ref{eq:exp_1}) and (\ref{eq:exp_2}), we have:
\begin{align*}
\bbL(Y, \hat{F})(t) - \bbL(Y, \hat{F}_0)(t) 
& \leq 2B * \bbC(Y, \hat{F})(t) + (B + 1)\epsilon_n
\end{align*}
Intergrating both sides of above equation with respect to $t \in \Ysc$, we have:
$$\bbL(Y, \hat{F}) - \bbL(Y, \hat{F}_0) \leq 2B * \bbC(Y, \hat{F}) + (B + 1)\epsilon_n$$
\end{proof}

Notice that Theorem \ref{thm:accuracy} is applicable to most of the standard measures for predictive accuracy, e.g. root mean squared error (RMSE). This is because for suitably standardized $\{y_i\}_{i=1}^N$, RMSE is always  bounded and proper \citep{gneiting_strictly_2007}. 

%% file: sections/table_app.tex
\begin{table}
\caption{Full results for the Calibration and Coverage probability metrics on out-of-sample data from the training time window $(-3\pi, 3\pi)$ and the test time window $(-5\pi, 5\pi)$ for the time-series regression task. 
\textbf{IND}: "IN-Domain" result from the training time window. \textbf{ALL}: full result from the entire test time window.
}
\label{tb:exp_1d_res_app}
\centering

\begin{adjustbox}{max width=0.95\linewidth,max totalheight=0.82\textheight}
\begin{tabular}{|c|c|c|c|c|c|c|c|c|c|}
\hline\hline
Metric & Model & $n=10$ & $n=25$ & $n=50$ & $n=100$ & $n=250$ & $n=500$ & $n=750$  & $n=1000$ \\ 
\hline
\multicolumn{10}{|c|}{Calibration Metrics}
\\
\hline
\multirow{5}{*}{NLL (IND)}
 &\textbf{Stacking} 
 & 0.7371 &  0.7191 & 0.6832 & 0.6682 & 0.6446 & 0.6326 & 0.6336 & 0.6428
\\
 &\textbf{GAM}
 & 1.0710 &  1.0877 & 1.1448 & 1.0068 & 1.2514 & 1.4000 & 1.5593 & 1.4739
\\
 &\textbf{BMA}
& 0.5056 & 0.5009 & 0.4846 & 0.4334 & 0.4615 & 0.4511 & 0.4439 & 0.4477
\\
 &\textbf{BAE}
& 0.2733 & 0.2672 & 0.2224 & 0.1752 & 0.2092 & 0.2062 & 0.2204 & 0.2099
\\
 &\textbf{BNE}
& \textbf{0.1105} & \textbf{0.0519} &\textbf{-0.0529} & \textbf{-0.1869} & \textbf{-0.3569} & \textbf{-0.4275} & \textbf{-0.4452} & \textbf{-0.4668}
\\ 
\hline
\multirow{5}{*}{NLL (ALL)}
 &\textbf{Stacking} 
 & 1.1658 & 1.1509 & 1.1281 & 1.0855 & 1.0434 & 1.0307 & 1.0437 & 1.0137
\\
 &\textbf{GAM}
 & 1.6492 & 1.7010 & 1.8538 & 1.6705 & 2.0532 & 2.2720 & 2.4573 & 2.3656
\\
 &\textbf{BMA}
& 0.8706 & 0.8508 & 0.8474 & 0.8048 & 0.8031 & 0.7761 & 0.7751 & 0.7814
\\
 &\textbf{BAE}
& 0.7489 & 0.7251 & 0.7030 & 0.6557 & 0.6603 & 0.6335 & 0.6417 & 0.6410
\\
 &\textbf{BNE}
& \textbf{0.5545} & \textbf{0.5103} & \textbf{0.4597} & \textbf{0.3853} & \textbf{0.2765} & \textbf{0.2386} & \textbf{0.2256} & \textbf{0.2238}
\\ 
\hline
\multirow{5}{*}{Calibration (IND)}
 &\textbf{Stacking} 
 & 0.1675 & 0.1682 & 0.1515 & 0.1521 & 0.1472 & 0.1470 & 0.1419 & 0.1462
\\
 &\textbf{GAM}
& \textbf{0.1478} & \textbf{0.1298} & \textbf{0.0928} & \textbf{0.0962} & \textbf{0.0542} & \textbf{0.0384} & \textbf{0.0333} & \textbf{0.0330} 
\\
 &\textbf{BMA}
& 0.5064 & 0.4087 & 0.3829 & 0.2923 & 0.2347 & 0.2078 & 0.1889 & 0.1913
\\
 &\textbf{BAE}
& 0.4282 & 0.3191 & 0.2504 & 0.1543 & 0.0862 & 0.0561 & 0.0465 & 0.0402
\\
 &\textbf{BNE}
& 0.2806 & 0.2677 & 0.2453 & 0.2004 & 0.1287 & 0.0873 & 0.0747 & 0.0621
\\ 
\hline
\multirow{5}{*}{Calibration (ALL)}
 &\textbf{Stacking} 
 & 0.5962 & 0.5999 & 0.5964 & 0.5697 & 0.5457 & 0.5449 & 0.5524 & 0.5138
\\
 &\textbf{GAM}
 & 0.7255 & 0.7429 & 0.7689 & 0.7566 & 0.7191 & 0.7925 & 0.8224 & 0.7671
\\
 &\textbf{BMA}
& 0.6148 & 0.5300 & 0.5153 & 0.4526 & 0.3906 & 0.3546 & 0.3321 & 0.3405
\\
 &\textbf{BAE}
& 0.6066 & 0.5118 & 0.4678 & 0.3958 & 0.3297 & \textbf{0.2862} & \textbf{0.2630} & \textbf{0.2687}
\\
 &\textbf{BNE}
& \textbf{0.3552} & \textbf{0.3513} & \textbf{0.3487} & \textbf{0.3448} & \textbf{0.3207} & 0.3150 & 0.2979 & 0.3055
\\ 
\hline
\multirow{5}{*}{Sharpness (IND)}
 &\textbf{Stacking} 
& 0.5696 & 0.5509 & 0.5317 & 0.5160 & 0.4974 & 0.4855 & 0.4916 & 0.4967 
\\
 &\textbf{GAM}
& 0.9231 & 0.9580  & 1.0520 & 0.9106 & 1.1972 & 1.3617 & 1.5260 & 1.4408
\\
 &\textbf{BMA}
& -0.0009 & 0.0922 & 0.1017 & 0.1411 & 0.2268 & 0.2433 & 0.2550 & 0.2565
\\
 &\textbf{BAE}
& -0.1549 & -0.0519 & -0.0279 & 0.0209 & 0.1230 & 0.1501 & 0.1740 & 0.1696
\\
 &\textbf{BNE}
& \textbf{-0.1701} & \textbf{-0.2157} & \textbf{-0.2982} & \textbf{-0.3873} & \textbf{-0.4856} & \textbf{-0.5149} & \textbf{-0.5198} & \textbf{-0.5289}
\\ 
\hline
\multirow{5}{*}{Sharpness (ALL)}
 &\textbf{Stacking} 
 & 0.5696 & 0.5510 & 0.5317 & 0.5158 & 0.4977 & 0.4858 & 0.4913 & 0.4966
\\
 &\textbf{GAM}
 & 0.9237 & 0.9580 & 1.0521 & 0.9103 & 1.1971 & 1.3619 & 1.5258 & 1.4409
\\
 &\textbf{BMA}
& 0.2558 & 0.3208 & 0.3321 & 0.3522 & 0.4125 & 0.4216 & 0.4430 & 0.4408
\\
 &\textbf{BAE}
& \textbf{0.1423} & 0.2133 & 0.2353 & 0.2599 & 0.3305 & 0.3473 & 0.3787 & 0.3723
\\
 &\textbf{BNE}
& 0.1994 & \textbf{0.1590} & \textbf{0.1111} & \textbf{0.0405} & \textbf{-0.0442} & \textbf{-0.0764} & \textbf{-0.0723} & \textbf{-0.0817}
\\ 
\hline
\hline
\multicolumn{10}{|c|}{Coverage Probability Metrics}
\\
\hline
\multirow{5}{*}{ECE (IND)} 
& \textbf{Stacking} &
0.0089 & 0.0084 &  0.0089 & 0.0086 & 0.0084 & 0.0083 & 0.0085 & 0.0084 
\\
 &\textbf{GAM}
& 0.0112 & 0.0115 & 0.0129 & 0.0123 & 0.0149 & 0.0161 & 0.0169 & 0.0168
\\
 &\textbf{BMA}
& 0.0065 & 0.0072 & 0.0073 & 0.0073 & 0.0079 & 0.0082 & 0.0080 & 0.0083
\\
 &\textbf{BAE}
& 0.0037 & 0.0043 & 0.0046 & 0.0045 & 0.0053 & 0.0057 & 0.0059 & 0.0059
\\
 &\textbf{BNE}
& \textbf{0.0021} & \textbf{0.0021} & \textbf{0.0019} & \textbf{0.0015} & \textbf{0.0013} & \textbf{0.0014} & \textbf{0.0015} & \textbf{0.0016}
\\ 
\hline
\multirow{5}{*}{ECE (ALL)} 
& \textbf{Stacking} 
& 0.0067 & 0.0063  & 0.0062 & 0.0063 & 0.0062 & 0.0060 & 0.0061 & 0.0064
\\
&\textbf{GAM}
& 0.0052 & 0.0048  & 0.0052 & 0.0049 & 0.0049 & 0.0042 & 0.0043 & 0.0045
\\
 &\textbf{BMA}
& 0.0064 & 0.0072 & 0.0070 & 0.0067 & 0.0075 & 0.0075 & 0.0075 & 0.0075
\\
 &\textbf{BAE}
& 0.0040 & 0.0046 & 0.0045 & 0.0042 & 0.0049 & 0.0051 & 0.0055 & 0.0052
\\
 &\textbf{BNE}
& \textbf{0.0023} & \textbf{0.0023} & \textbf{0.0020} & \textbf{0.0017} & \textbf{0.0013} & \textbf{0.0014} & \textbf{0.0015} & \textbf{0.0014}
\\ 
\hline
\multirow{5}{*}{\makecell{$95\%$ CI \\ Coverage \\ Probability (IND)}}
& \textbf{Stacking}
& 0.9671 & 0.9679 & 0.9689 & 0.9669 & 0.9666 & 0.9671 & 0.9669 &
0.9666 \\
&\textbf{GAM}
& 0.9730 & 0.9789 & 0.9805 & 0.9806 & 0.9825 & 0.9851 & 0.9876 &
0.9859 \\
 &\textbf{BMA}
& 0.9129 & 0.9265 & 0.9303 & 0.9397 & 0.9478 & 0.9508 & 0.9532 & 0.9527
\\
 &\textbf{BAE}
& 0.8996 & 0.9158 & 0.9227 & 0.9339 & 0.9494 & 0.9564 & 0.9583 & 0.9583
\\
 &\textbf{BNE}
& 0.9257 & 0.9258 & 0.9305 & 0.9312 & 0.9471 & 0.9562 & 0.9592 & 0.9615
\\ 
\hline
\multirow{5}{*}{\makecell{$95\%$ CI \\ Coverage \\ Probability (ALL)}}
& \textbf{Stacking} 
& 0.9283 & 0.9279 & 0.9277 & 0.9297 & 0.9314 & 0.9337 & 0.9314 & 0.9347 \\
&\textbf{GAM}
& 0.9139 & 0.9121 & 0.9056 & 0.9094 & 0.9086 & 0.8995 & 0.8963 & 0.9032 \\
 &\textbf{BMA}
& 0.9178 & 0.9284 & 0.9322 & 0.9382 & 0.9461 & 0.9496 & 0.9523 & 0.9510
\\
 &\textbf{BAE}
 & 0.9066 & 0.9184 & 0.9241 & 0.9321 & 0.9441 & 0.9503 & 0.9533 & 0.9521
\\
 &\textbf{BNE}
 & 0.9460 & 0.9452 & 0.9475 & 0.9443 & 0.9511 & 0.9519 & 0.9544 & 0.9546
\\ 
\hline
\end{tabular}
\end{adjustbox}
\end{table}

\newpage

\begin{table}
\caption{Full results for the Calibration and Coverage probability metrics on out-of-sample data from the training region $(-\pi, \pi)^2$ and the test region $(-1.25\pi, 1.25\pi)^2$ for the spatial regression task. \textbf{IND}: "IN-Domain" result from the training region. \textbf{ALL}: full result from the entire test region.
}
\label{tb:exp_2d_res_app}
\centering

\begin{adjustbox}{max width=0.95\linewidth,max totalheight=0.82\textheight}
\begin{tabular}{|c|c|c|c|c|c|c|c|c|}
\hline\hline
Metric & Model & $n=25$ & $n=50$ & $n=100$ & $n=250$ & $n=500$ & $n=750$  & $n=1000$ \\ 
\hline
\multicolumn{9}{|c|}{Calibration Metrics}
\\
\hline
\multirow{5}{*}{NLL (IND)}
 &\textbf{Stacking} 
 & -2.2334  & -2.2524 & -2.2594 & -2.2655 & -2.2713 & -2.2691 & -2.2772 
\\
 &\textbf{GAM}
 & -1.9349 & -2.1350 & -2.2897 & -2.5740 & -2.7646 & -2.8114 & -2.8071
\\
 &\textbf{BMA}
& -2.8318 & -2.8464 & -2.8708 & -2.9216 & -2.9458 & -2.9679 & -2.9731
\\
 &\textbf{BAE}
& -3.0345 & -3.0525 & -3.0711 & -3.1191 & -3.1381 & -3.1564 & -3.1649
\\
 &\textbf{BNE}
& -3.1267 & -3.1407 & -3.1648 & -3.2149 & -3.2457 & -3.2569 & -3.2695
\\ 
\hline
\multirow{5}{*}{NLL (ALL)}
 &\textbf{Stacking} 
 & -1.7356 & -1.7605  & -1.7815 & -1.7994 & -1.8106 & -1.8151 & -1.8188
\\
 &\textbf{GAM}
 & -1.0028 & -1.1653 & -1.2733 & -1.5408 & -1.7205 & -1.7676 & -1.7639
\\
 &\textbf{BMA}
& -2.1288 & -2.1632 & -2.1988 & -2.2587 & -2.2914 & -2.3066 & -2.2892
\\
 &\textbf{BAE}
& -2.4281 & -2.4555 & -2.4836 & -2.5292 & -2.5582 & -2.5720 & -2.5694
\\
 &\textbf{BNE}
& -2.5229 & -2.5466 & -2.5772 & -2.6235 & -2.6643 & -2.6799 & -2.6841
\\ 
\hline
\multirow{5}{*}{Calibration (IND)}
 &\textbf{Stacking} 
  & 0.5851 & 0.5907 & 0.6052 & 0.6185 & 0.6252 & 0.6331 & 0.6233
\\
 &\textbf{GAM}
 & 0.1468 & 0.1131 & 0.0688 & 0.0461 & 0.0411 & 0.0398 & 0.0380
\\
 &\textbf{BMA}
& 0.7644 & 0.7578 & 0.7406 & 0.7063 & 0.6993 & 0.6872 & 0.6853
\\
 &\textbf{BAE}
& 0.6560 & 0.6462 & 0.6368 & 0.6080 & 0.6065 & 0.5966 & 0.5917
\\
 &\textbf{BNE}
& 0.5917 & 0.5864 & 0.5724 & 0.5445 & 0.5359 & 0.5355 & 0.5298
\\ 
\hline
\multirow{5}{*}{Calibration (ALL)}
 &\textbf{Stacking} 
 & 1.0823 & 1.0822  & 1.0833 & 1.0847 & 1.0857 & 1.0869  & 1.0819
\\
 &\textbf{GAM}
 & 1.0787 &  1.0828 & 1.0854 & 1.0796 & 1.0852 & 1.0831 & 1.0812 
\\
 &\textbf{BMA}
&  1.0430 & 1.0210 & 1.0000 & 0.9682 & 0.9683 & 0.9692 & 0.9888
\\
&\textbf{BAE}
& 0.8098 & 0.7952 & 0.7834 & 0.7687 & 0.7740 & 0.7754 & 0.7807
\\
 &\textbf{BNE}
& 0.6964 & 0.6854 & 0.6698 & 0.6530 & 0.6448 & 0.6450 & 0.6444
\\ 
\hline
\multirow{5}{*}{Sharpness (IND)}
 &\textbf{Stacking} 
& -2.8185 & -2.8431  & -2.8646 & -2.8840 & -2.8965 & -2.9022 & -2.9005
\\
 &\textbf{GAM}
& -2.0818 & -2.2481 & -2.3586 & -2.6200 & -2.8057 & -2.8511 & -2.8450
\\
 &\textbf{BMA}
& -3.5962 & -3.6042 & -3.6115 & -3.6279 & -3.6450 & -3.6551 & -3.6584
\\
 &\textbf{BAE}
& -3.6905 & -3.6988 & -3.7079 & -3.7271 & -3.7446 & -3.7529 & -3.7566
\\
 &\textbf{BNE}
& -3.7184 & -3.7271 & -3.7372 & -3.7594 & -3.7816 & -3.7925 & -3.7993
\\ 
\hline
\multirow{5}{*}{Sharpness (ALL)}
 &\textbf{Stacking} 
 & -2.8185 & -2.8431 & -2.8646 & -2.8840 & -2.8965 & -2.9022 & 
-2.9005 \\
 &\textbf{GAM}
 & -2.0818 & -2.2481 & -2.3586 & -2.6200 & -2.8057 & -2.8511 & -2.8450 \\
 &\textbf{BMA}
& -3.1719 & -3.1842 & -3.1988 & -3.2269 & -3.2597 & -3.2758 & -3.2780
\\
 &\textbf{BAE}
& -3.2379 & -3.2507 & -3.2671 & -3.2980 & -3.3323 & -3.3474 & -3.3501
\\
 &\textbf{BNE}
& -3.2192 & -3.2319 & -3.2470 & -3.2765 & -3.3091 & -3.3249 & -3.3286
\\ 
\hline
\hline
\multicolumn{9}{|c|}{Coverage Probability Metrics}
\\
\hline
\multirow{5}{*}{ECE (IND)} 
& \textbf{Stacking} 
& 0.0042 & 0.0040  & 0.0039 & 0.0037 & 0.0036 & 0.0035 & 0.0036
\\
&\textbf{GAM}
& 0.0162 & 0.0173  & 0.0189 & 0.0199 & 0.0202 & 0.0203 & 0.0203
\\
 &\textbf{BMA}
& 0.0056 & 0.0058 & 0.0059 & 0.0060 & 0.0061 & 0.0061 & 0.0062
\\
 &\textbf{BAE}
& 0.0043 & 0.0043 & 0.0043 & 0.0045 & 0.0046 & 0.0047 & 0.0049
\\
 &\textbf{BNE}
& 0.0048 & 0.0048 & 0.0049 & 0.0051 & 0.0051 & 0.0052 & 0.0054
\\ 
\hline
\multirow{5}{*}{ECE (ALL)} 
& \textbf{Stacking} 
& 0.0042 & 0.0040  & 0.0039 & 0.0037 & 0.0036 & 0.0035 & 0.0036
\\
 &\textbf{GAM}
& 0.0162 & 0.0147  & 0.0173 & 0.0189 & 0.0199 & 0.0202 & 0.0203
\\
 &\textbf{BMA}
& 0.0132 & 0.0131 & 0.0125 & 0.0121 & 0.0115 & 0.0114 & 0.0117
\\
 &\textbf{BAE}
& 0.0056 & 0.0054 & 0.0051 & 0.0046 & 0.0043 & 0.0041 & 0.0042
\\
 &\textbf{BNE}
& 0.0044 & 0.0042 & 0.0039 & 0.0036 & 0.0034 & 0.0033 & 0.0035
\\ 
\hline
\multirow{5}{*}{\makecell{$95\%$ CI \\ Coverage \\ Probability (IND)}}
& \textbf{Stacking} 
& 0.9648 & 0.9639  & 0.9597 & 0.9581 & 0.9566 & 0.9558 & 0.9568
\\
&\textbf{GAM}
& 0.9957 & 0.9973  & 0.9993 & 0.9998 & 0.9999 & 1.0000 & 1.0000 
\\
 &\textbf{BMA}
& 0.9400 & 0.9390 & 0.9429 & 0.9445 & 0.9464 & 0.9443 & 0.9462
\\
 &\textbf{BAE}
& 0.9449 & 0.9454 & 0.9473 & 0.9507 & 0.9520 & 0.9517 & 0.9519
\\
 &\textbf{BNE}
& 0.9570 & 0.9577 & 0.9579 & 0.9616 & 0.9641 & 0.9618 & 0.9629
\\ 
\hline
\multirow{5}{*}{\makecell{$95\%$ CI \\ Coverage \\ Probability (ALL)}}
& \textbf{Stacking} 
& 0.9083 & 0.9080  & 0.9060 & 0.9060 & 0.9059 & 0.9056 & 0.9066
\\
&\textbf{GAM}
& 0.9253  & 0.9288 & 0.9303 & 0.9335 & 0.9305 & 0.9297 & 0.9287 
\\
 &\textbf{BMA}
 & 0.8486 & 0.8491 & 0.8536 & 0.8561 & 0.8586 & 0.8568 & 0.8574
\\
 &\textbf{BAE}
 & 0.8762 & 0.8777 & 0.8792 & 0.8827 & 0.8846 & 0.8848 & 0.8848
\\
 &\textbf{BNE}
 & 0.9298 & 0.9308 & 0.9325 & 0.9336 & 0.9360 & 0.9340 & 0.9336
\\ 
\hline
\end{tabular}
\end{adjustbox}
\end{table}

\clearpage

\begin{table}[ht]
\caption{Mean and Standard Deviation for spatial 6-fold cross validation \glsfirst{RMSE}, \glsfirst{NLL} and coverage probability of $95\%$ \glsfirst{CI}
for 2011 PM$_{2.5}$ predictions in Eastern New England, USA. (Full Results)}
\label{tb:app_comp_full}
\centering
\begin{adjustbox}{max width=\linewidth}
\begin{tabular}{|c|c|c|c|c|c|}
\hline\hline

\multirow{3}{*}{Model/Metric} & \multirow{3}{*}{\makecell{\textbf{RMSE}(mean)}} & \multirow{3}{*}{\textbf{RMSE}(median)}  & \multirow{3}{*}{\textbf{NLL}(mean)}  & \multirow{3}{*}{\textbf{NLL}(median)} & 
\makecell{$95\%$ CI \\ Coverage \\ Probability}
 \\  \hline

\textbf{Stacking} & $0.862 \pm 0.292$ & $0.785 \pm 0.292$ & $0.941 \pm 0.088$ & $0.914 \pm 0.088$ & $0.9412$ \\ 
\hline 
\textbf{GAM} & $1.096 \pm 0.161$ & $1.116 \pm 0.161$ & $1.208 \pm  0.409$ & $1.008 \pm 0.409$ & $0.9216$ \\
\hline 
\textbf{BMA} & $0.888 \pm 0.423$ & $0.842 \pm 0.423$ & $1.081 \pm  0.345$ & $0.930 \pm 0.345$ & $0.9412$ \\
\hline 
\makecell{\textbf{BAE}  \\ \scriptsize{(Prior Mean=Zero)}} & $ 0.872 \pm 0.397$ & $0.767 \pm 0.397$ & $1.051 \pm  0.223$ & $0.905 \pm 0.223$ & $0.9412$ \\
\hline 
\makecell{\textbf{BAE} \\ \scriptsize{(Prior Mean=Stacking)}} & $0.841 \pm 0.348$ & $0.719 \pm 0.348$  & $0.967 \pm 0.137$ & $0.897 \pm 0.137$ & $0.9412$ \\ 
\hline 
\makecell{\textbf{BAE} \\ \scriptsize{(Prior Mean=GAM)}} & $0.882 \pm 0.234$ & $0.827 \pm 0.234$  & $1.116 \pm 0.410$ & $ 0.929 \pm 0.410$ & $0.9216$ \\ 
\hline 
\makecell{\textbf{BAE} \\ \scriptsize{(Prior Mean=Stacking+GAM)}} & $0.834 \pm 0.297$ & $0.728 \pm 0.297$  & $1.039 \pm 0.296$ & $0.927 \pm 0.296$ & $0.8824$ \\ 
\hline 
\makecell{\textbf{BNE} \scriptsize{(variance only)} \\ \scriptsize{(Prior Mean=Zero)}} & $0.872 \pm 0.397$ & $0.775 \pm 0.397$  & $1.053 \pm  0.212$ & $0.928 \pm  0.212$ & $0.9412$\\ 
\hline  
\makecell{\textbf{BNE} \scriptsize{(variance only)}\\ \scriptsize{(Prior Mean=Stacking)}} & $0.840 \pm 0.349$ & $0.724 \pm 0.349$  & $0.962 \pm  0.127$ & $0.892 \pm  0.127$ & $0.9608$\\
\hline
\makecell{\textbf{BNE} \scriptsize{(variance only)} \\ \scriptsize{(Prior Mean=GAM)}} & $ 0.879 \pm 0.238$ & $ 0.828 \pm 0.238$  & $1.096 \pm 0.377$ & $0.916 \pm 0.377$ & $0.9216$ \\ 
\hline 
\makecell{\textbf{BNE} \scriptsize{(variance only)}\\ \scriptsize{(Prior Mean=Stacking+GAM)}} & $0.832 \pm 0.300$ & $0.731 \pm 0.300$  & $1.024 \pm 0.270$ & $0.922 \pm 0.270$ & $0.9020$ \\
\hline 
\makecell{\textbf{BNE} \scriptsize{(variance+skewness)} \\ \scriptsize{(Prior Mean=Zero)}} & $0.853 \pm 0.391$ & $0.755 \pm 0.391$  & $1.091 \pm  0.278$ & $0.910 \pm  0.278$ & $0.9412$\\ 
\hline  
\makecell{\textbf{BNE} \scriptsize{(variance+skewness)} \\ \scriptsize{(Prior Mean=Stacking)}} & $0.831 \pm 0.348$ & $0.715 \pm 0.348$  & $0.972 \pm  0.137$ & $0.894 \pm  0.137$ & $0.9412$\\
\hline
\makecell{\textbf{BNE} \scriptsize{(variance+skewness)}\\ \scriptsize{(Prior Mean=GAM)}} & $0.871 \pm 0.236$ & $0.809 \pm 0.236$  & $1.098 \pm  0.373$ & $0.922 \pm  0.373$ & $0.9216$ \\ 
\hline 
\makecell{\textbf{BNE} \scriptsize{(variance+skewness)}\\ \scriptsize{(Prior Mean=Stacking+GAM)}} & $0.823 \pm 0.298$ & $0.716 \pm 0.298$  & $1.035 \pm 0.283$ & $0.931 \pm 0.283$ & $0.9020$ \\ 
\hline 
\hline 

\end{tabular}
\end{adjustbox}
\end{table}

%% file: sections/figure_app.tex

  


\begin{figure}[ht]
    \centering
    \includegraphics[width=.32\linewidth]{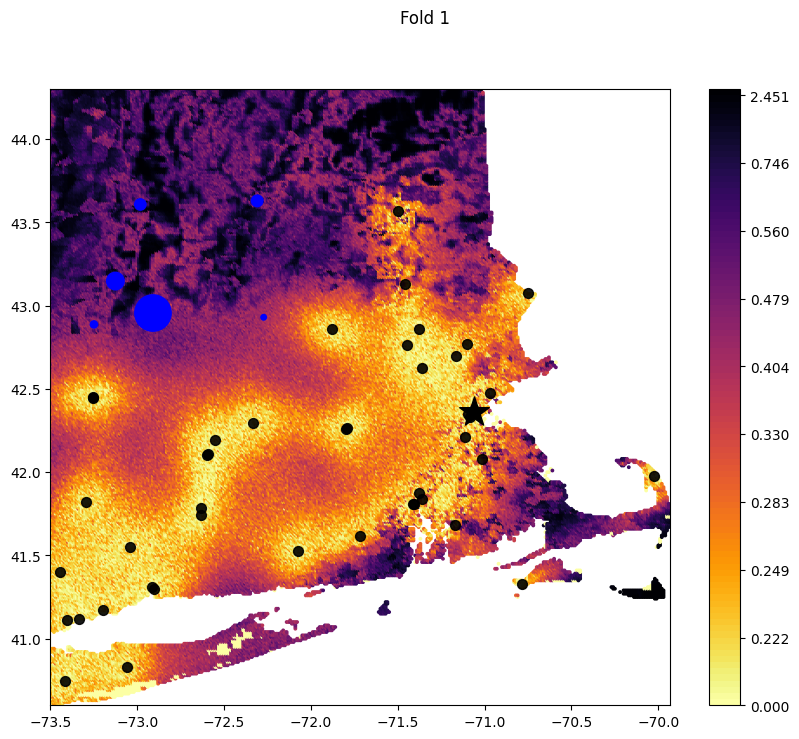}
    \includegraphics[width=.32\linewidth]{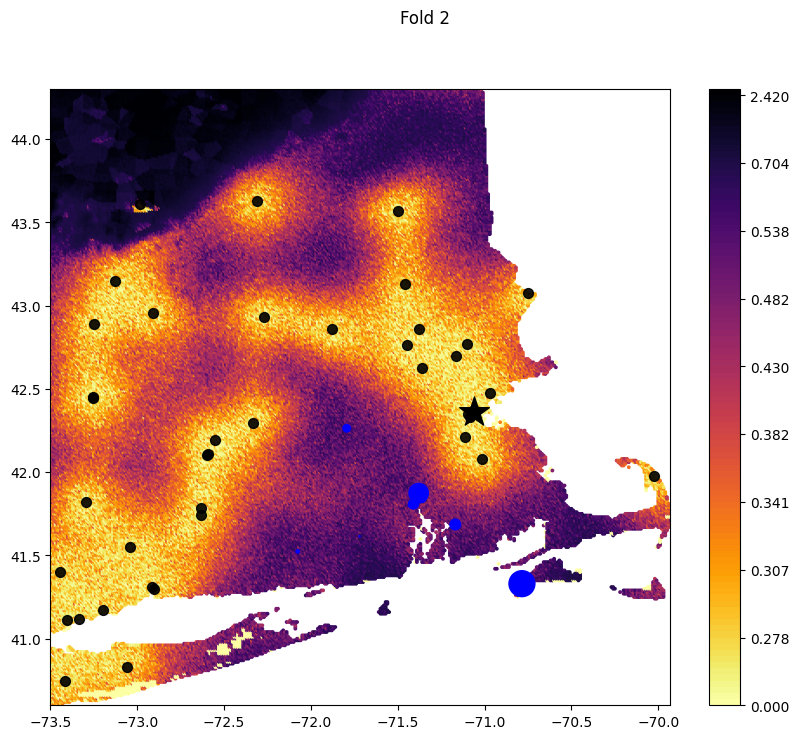}
    \includegraphics[width=.32\linewidth]{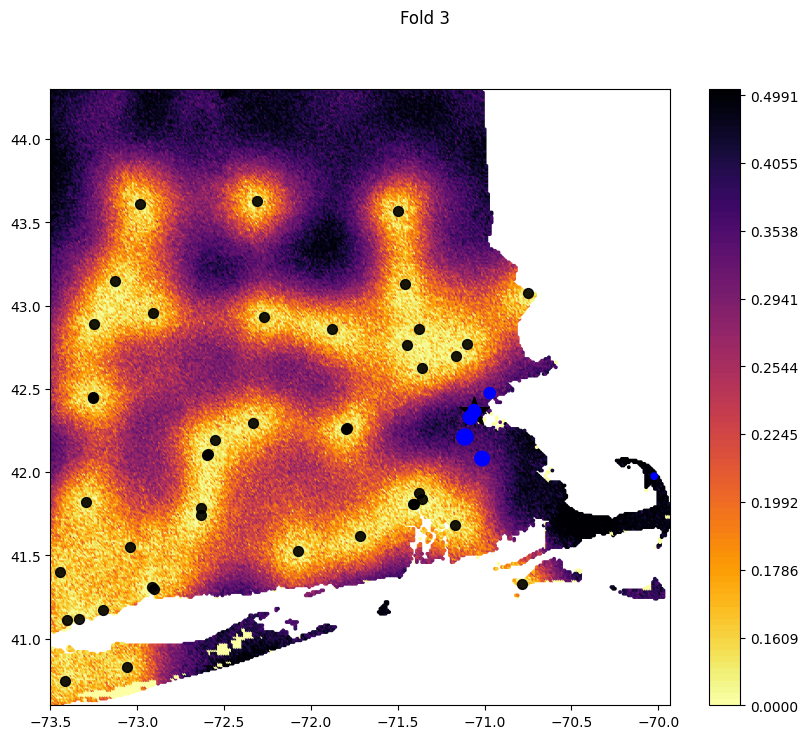}
    
    \includegraphics[width=.32\linewidth]{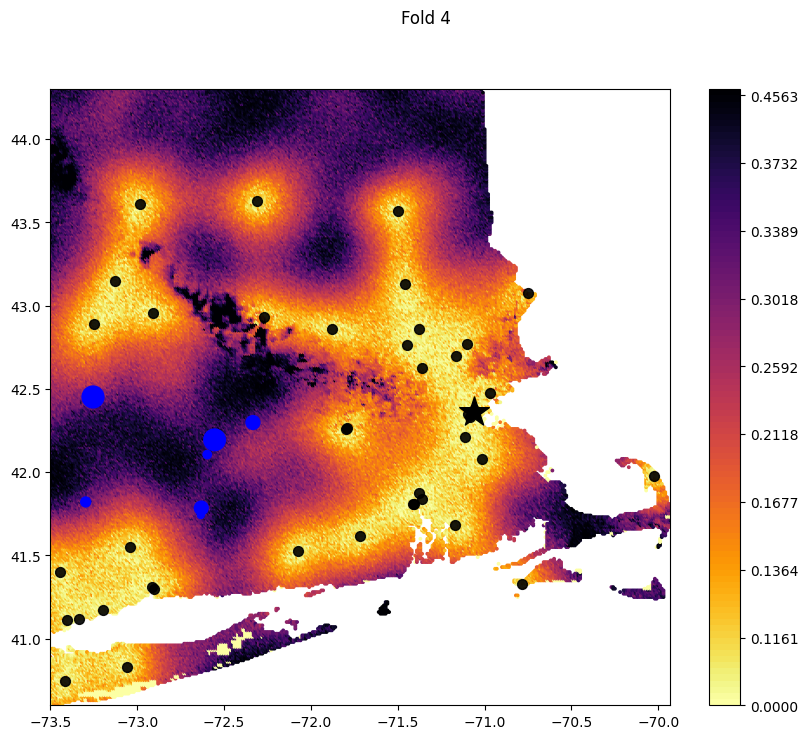}
    \includegraphics[width=.32\linewidth]{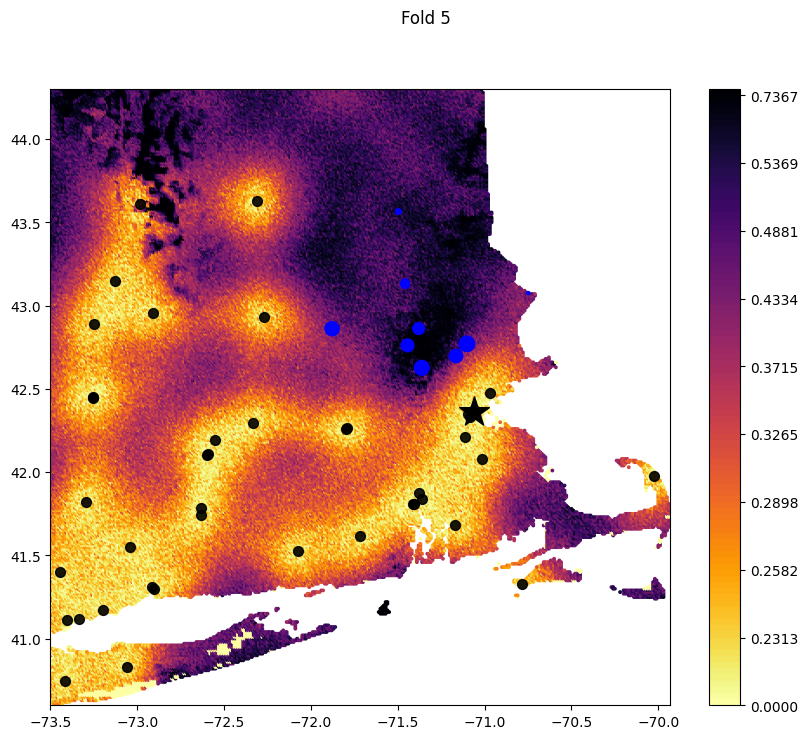}
    \includegraphics[width=.32\linewidth]{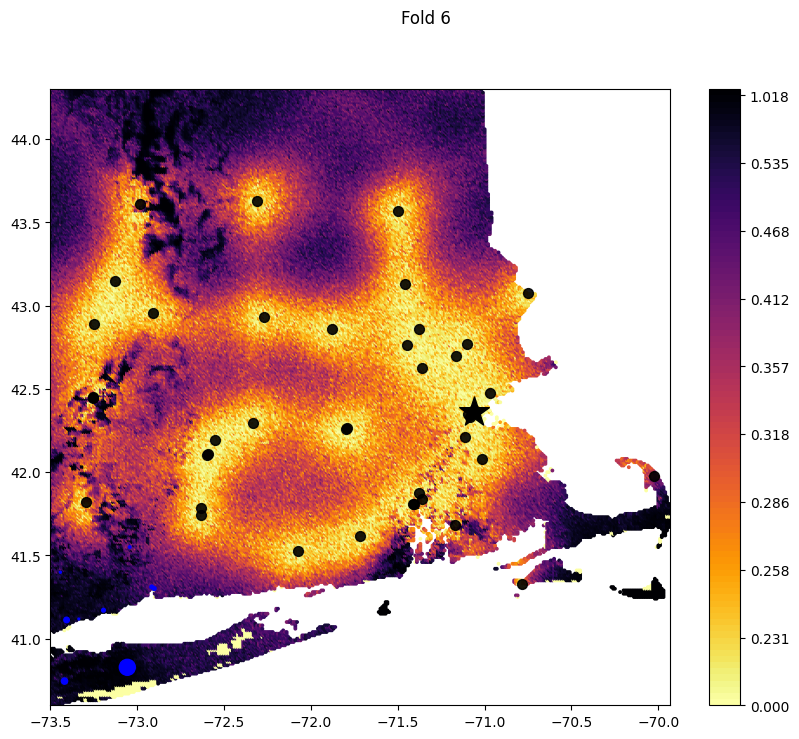}
    \caption{
    Spatial Cross Validation BAE plots for \PM  prediction in Eastern Massachusetts
    \textbf{Blue Dots}: Test Monitor Locations; \textbf{Black Dots}: Train Monitor Locations; \textbf{Black Star}: Boston, MA.}
    \label{fig:spcv}
\end{figure}